\RequirePackage{silence}
\documentclass{aa}  
\usepackage{graphicx}
\usepackage{txfonts}
\usepackage{url}

\usepackage{booktabs}
\usepackage{siunitx}
\usepackage{enumitem} 
\setlist[enumerate]{itemsep=1em} 
\usepackage{float}
\usepackage{placeins}
\usepackage{tabularx}
\usepackage{rotating}
\usepackage{longtable}
\usepackage{threeparttablex} 
\usepackage{pdflscape}
\usepackage{xcolor}
\usepackage[colorlinks=true, linkcolor=blue, citecolor=blue, urlcolor=cyan, hypertexnames=false]{hyperref}

\begin{document}

   \title{Radial dust distributions and obscuring geometry in AGN from JWST/MIRI spectroscopy}

\author{
  Ruiyu Pan\inst{1,2}
  \and
  Arkaprabha Sarangi\inst{3,4}
}

\institute{
  Niels Bohr Institute, University of Copenhagen, Lyngbyvej 2, 2100 Copenhagen Ø, Denmark
  \and
  School of Physics and Astronomy, University of Birmingham, Edgbaston, Birmingham B15 2TT, UK
  \and
  Indian Institute of Astrophysics, 100 Feet Road, Koramangala, Bengaluru, Karnataka 560034, India
  \and
  DARK, Niels Bohr Institute, University of Copenhagen, Jagtvej 155A, 2200 Copenhagen, Denmark
}

   \date{}

\abstract{The spatial distribution of obscuring dust in active galactic nuclei (AGN) is critical for distinguishing between static torus models and dynamical disk-wind scenarios. To constrain this geometry, we forward-model the rest-frame $5$--$14\,\mu\mathrm{m}$ JWST/MIRI MRS spectra of 25 local AGN using a three-dimensional, axisymmetric radiative-transfer library combined with empirical starburst templates. Using a grid-based inference framework, we systematically compare radial dust-density laws of the form $n(r)\propto r^{-p}$ over the range $p=0.5$--$2.0$. Our model comparison strongly favours shallow radial profiles at the sample level: 20 sources achieve their largest statistical weight at $p=0.5$, and 21 accumulate more than half of their combined density-law weight at $p\leq1$. This tendency remains robust even when the likelihood power $\beta$ is varied, although the preferred profile of individual sources can change. Furthermore, the $9.7\,\mu\mathrm{m}$ silicate feature exhibits a distinct trend: absorption minima remain near $9.7\,\mu\mathrm{m}$, whereas four observed emission maxima are shifted redward by approximately $0.8$--$1.3\,\mu\mathrm{m}$. These results favour relatively extended MIR-emitting dust distributions rather than strictly compact geometries, and suggest that an interplay of radiative-transfer effects and intrinsic dust grain properties drives the observed spectral diversity.}


   \maketitle
%
%-------------------------------------------------------------------
\section{Introduction}

Dust surrounding AGN absorbs radiation from the accretion disk and re-emits a substantial fraction of it at infrared wavelengths. In the classical orientation-based unification framework, the observed distinction between type~1 and type~2 AGN is attributed primarily to obscuration by an optically thick circumnuclear dust distribution \citep{Antonucci1985,Antonucci1993,Pier1992,2015ARA&A..53..365N}. The structure responsible for this obscuration has traditionally been described as a geometrically thick torus. However, the physical origin, spatial distribution, and dynamical support of this dusty medium remain uncertain.

High-angular-resolution infrared observations have substantially revised the classical picture of a static, smooth torus. Interferometric observations resolve compact equatorial components together with extended emission that is frequently elongated along the polar direction \citep{2004Natur.429...47J,2012ApJ...755..149H,Lopez-Gonzaga2016}. Subarcsecond MIR imaging further indicates that polar dust emission is common in nearby Seyfert galaxies \citep{2016ApJ...822..109A,2019MNRAS.489.2177A}. These observations suggest that a significant fraction of the AGN-heated dust is associated with a dynamical disk--outflow structure rather than being confined to a purely equatorial torus.

Several physical mechanisms have been proposed to generate and maintain such geometrically and radially extended obscuring structures. Magnetocentrifugal disk winds can lift material from the accretion disk along ordered magnetic-field lines \citep{1982MNRAS.199..883B,2006ApJ...648L.101E}, while radiation pressure acting on dust can produce polar outflows and circulation between inflowing and outflowing gas \citep{2012ApJ...758...66W}. Radiation-magnetohydrodynamic simulations similarly generate geometrically thick, multiphase structures containing an equatorial gas reservoir and lower-density polar outflows \citep{2017ApJ...843...58C,2019ApJ...876..137W,2020ApJ...897...26W}. Radiative-transfer calculations of disk--wind geometries show that these structures can reproduce many of the observed infrared properties of AGN \citep{2017ApJ...838L..20H}.

One of the uncertainties in these models is the radial distribution of the dusty material. Smooth, clumpy, and two-phase torus models commonly parameterise the radial density as a power law, $n(r)\propto r^{-p}$ \citep{Fritz2006,2008ApJ...685..160N,2010A&A...523A..27H,Stalevski_2012}. The density index controls the relative contributions of hot dust near the sublimation front and cooler material at larger radii, thereby affecting the MIR continuum slope, silicate-feature profile, and apparent spatial extent of the emission. Self-similar magnetohydrodynamic (MHD) disk-wind models often produce density distributions close to $n(r)\propto r^{-1}$ \citep{Contopoulos1994,Fukumura2010,Behar2009,Sarangi_2019}, but different mass-loading, acceleration, and geometrical configurations can yield shallower or steeper effective profiles. It therefore remains unclear whether a single radial density law can describe the diversity of observed AGN MIR spectra.

The silicate features near $9.7$ and $18\,\mu\mathrm{m}$ provide an additional diagnostic of the circumnuclear dust. Their observed strengths and profiles depend on viewing geometry, optical depth, temperature gradients, and the relative contributions of directly illuminated and obscured dust \citep{2009ApJ...707.1550N,2015ApJ...803..110H}. In particular, the maximum of the silicate emission feature is frequently observed at wavelengths longer than $9.7\,\mu\mathrm{m}$. Radiative-transfer effects can produce part of this displacement, while variations in grain size, porosity, and composition can further modify the feature profile \citep{2008MNRAS.391L..49L,2017ApJS..228....6X,2023A&A...676A..73G}. Measurements of both the strength and peak wavelength of the silicate feature therefore provide complementary constraints on the geometry and dust properties of the emitting region.

Interpreting unresolved MIR spectra also requires separating the AGN-heated dust emission from circumnuclear star formation. Polycyclic aromatic hydrocarbon (PAH) features and cooler dust continua can contribute significantly to the integrated spectrum, particularly in composite and infrared-luminous systems \citep{Genzel1998,Laurent2000,Spoon2007}. Spectral-decomposition methods consequently combine AGN continua with empirical or theoretical star-forming templates to estimate their relative contributions \citep{Nardini2008,2006ApJ...653.1129B,2007ApJ...656..770S}. The inferred AGN fraction may nevertheless remain sensitive to the assumed intrinsic torus spectrum, and hence to the adopted radial density distribution.

The wavelength coverage and sensitivity of the JWST MIRI now make it possible to examine these effects using spatially resolved or nuclear MIR spectra with substantially improved spectral fidelity. Recent MIRI observations reveal considerable diversity in the continua, silicate features, and star-formation signatures of nearby AGN and type~2 quasars \citep{2025MNRAS.539.2158G,2025AA...698A.194R,2025arXiv251202629R}. These data provide a suitable basis for testing whether variations in MIR spectral shape can be explained by viewing geometry alone or require intrinsic differences in the radial dust distribution.

In this work, we develop a three-dimensional, axisymmetric radiative-transfer framework for AGN-heated circumnuclear dust. The density field is based on an MHD-motivated disk-wind geometry, while its radial dependence is parameterised as $n(r,\theta)\propto r^{-p}$. We construct model libraries for $p=\{0.5,1.0,1.5,2.0\}$ and apply them to rest-frame $5$--$14\,\mu\mathrm{m}$ spectra obtained with the MRS of JWST/MIRI for 25 AGN. The torus and empirical starburst templates are fitted to the spectra over the full model grid. We compare the four radial density laws, examine the silicate emission and absorption profiles, and determine what the spectra imply for the other model parameters.

The paper is organised as follows. Section~\ref{2} describes the physical assumptions, radiative-transfer calculations, and spectral-fitting framework. Section~\ref{sec:sample} introduces the sample and observational data. Section~\ref{3} presents the model spectra and spectral-decomposition results. Section~\ref{discussion} discusses the physical implications of the inferred density distributions, silicate-feature behaviour, and obscuring geometry, and summarises the main conclusions.

\section{Model}
\label{2}
\subsection{AGN}

\begin{figure}[t]
\centering
\includegraphics[width=\linewidth]{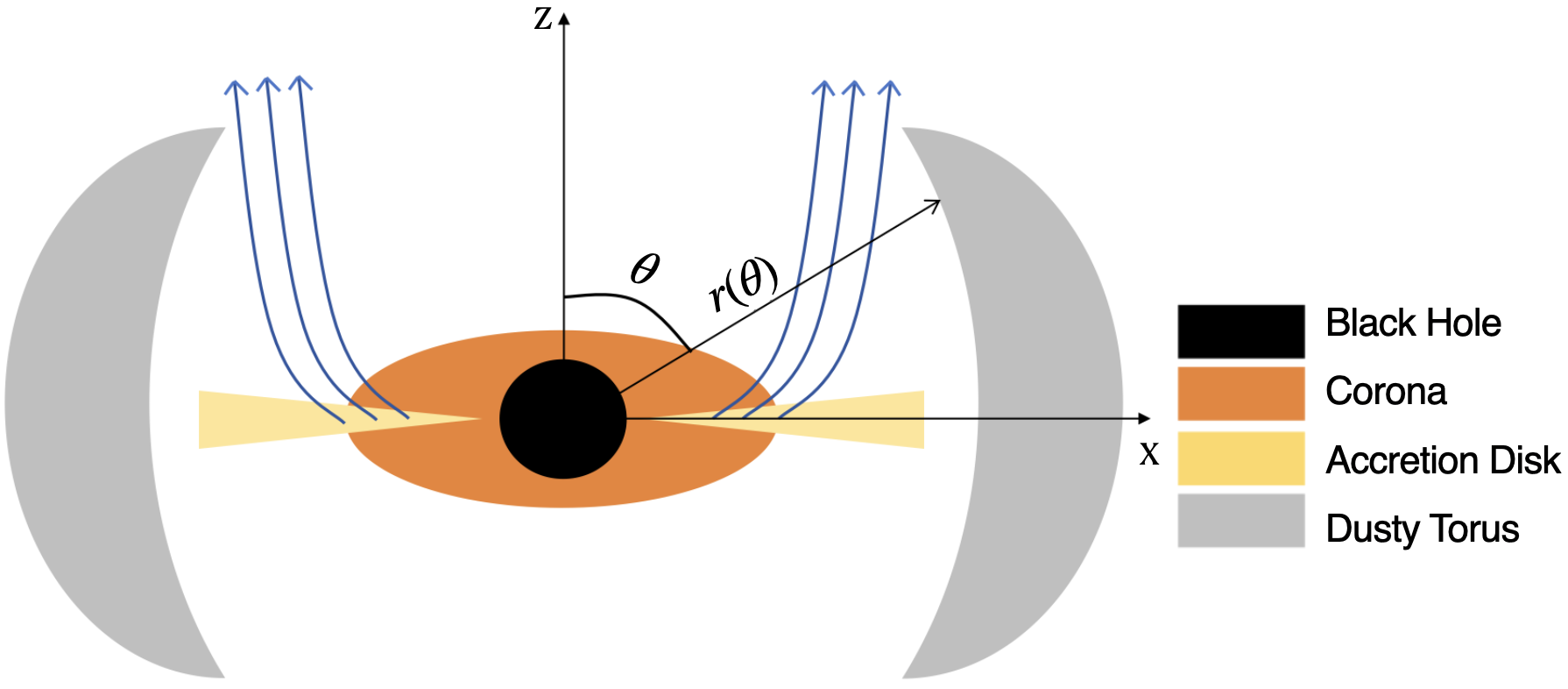}
\caption{Simplified 2D schematic of an AGN, illustrating the central black hole, hot corona, accretion disk, dusty torus, and disk wind. The quantity $r(\theta)$ denotes the angle-dependent inner radius of the dusty region.}
\label{AGNtructure}
\end{figure}

\begin{figure*}[t]
\centering
\includegraphics[width=\linewidth]
{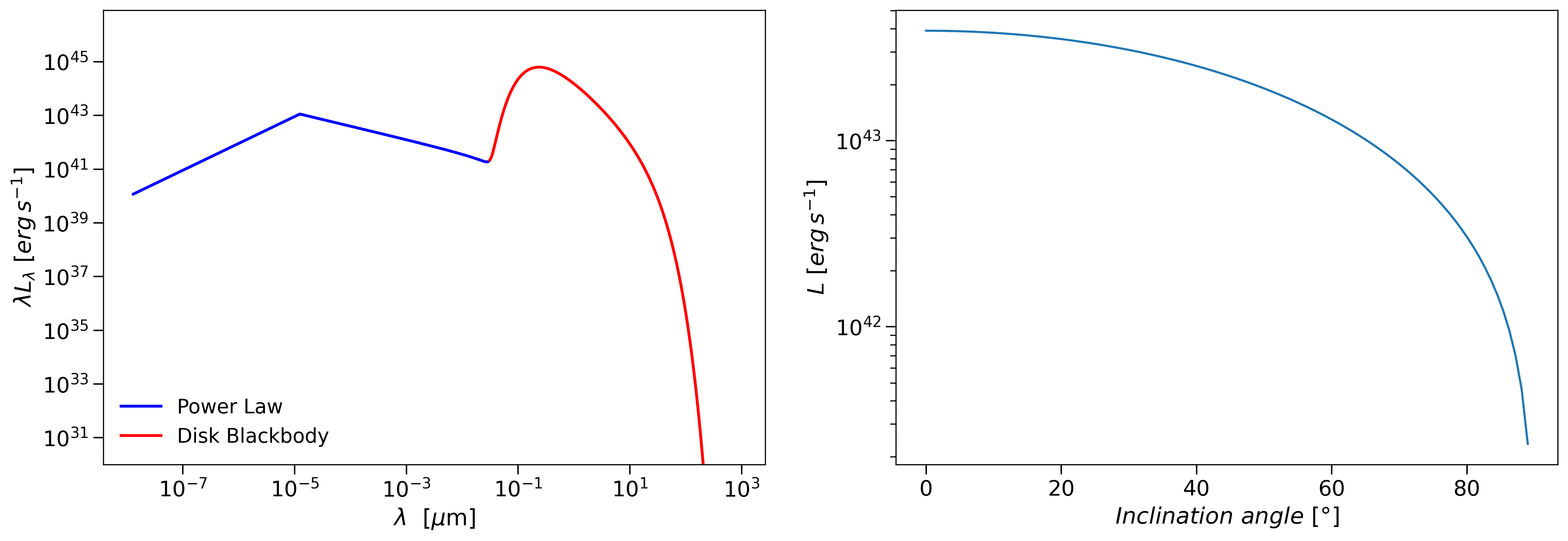}
\caption{Left panel: spectral energy distribution of the central AGN source at an inclination angle of $\theta = 75^\circ$. The spectral shape, characterised by $\alpha_{\mathrm{ox}} = -1.5$, remains consistent across all viewing angles \citep{Sarangi_2019}. The surrounding material is entirely heated by the central AGN radiation.
Right panel: angular luminosity modulation, following the luminosity profile defined in Eq.~(\ref{eq:ftheta}), showing how the directional luminosity normalisation varies with inclination.}
\label{AGNpectra}
\end{figure*}

We model the AGN as a central supermassive black hole surrounded by a geometrically thin, optically thick accretion disk and an X-ray emitting hot corona (Fig.~\ref{AGNtructure}). The disk emission is described by a thermal component, while the coronal continuum is represented by a power law produced by inverse-Compton scattering of disk photons \citep{Shakura1973,Haardt1993}. The balance between these two components, often characterised by the optical-to-X-ray spectral index ($\alpha_{\rm OX}$), determines the overall shape of the continuum \citep{Lusso2016}. The resulting spectral energy distribution (SED) shape is assumed to be independent of viewing angle (Fig.~\ref{AGNpectra}; \citealt{Sarangi_2019}); only the directional luminosity normalisation varies with viewing angle.

To account for anisotropic emission from a disk-like source, we scale the emitted luminosity with polar angle $\theta$ as
\begin{equation}
L(\theta)=\eta\,\dot{m}\,L_{\rm Edd}(M_8)\,f(\theta),
\label{eq:Ltheta}
\end{equation}
where $\eta$ denotes the adopted radiative efficiency, and $\dot{m} = \dot{M}/\dot{M}_{\mathrm{Edd}}$ is the normalised accretion rate. The Eddington accretion rate is defined as $\dot{M}_{\mathrm{Edd}} = L_{\mathrm{Edd}} / c^2$, with $L_{\mathrm{Edd}}$ representing the Eddington luminosity, which scales linearly with the black hole mass $M_{\rm BH}$. For convenience, the black hole mass is expressed in units of $M_8 = M_{\rm BH} / 10^8M_\odot$ \citep{Sarangi_2019}. The angular modulation is described by \citep{1987MNRAS.225...55N,Stalevski_2012}
\begin{equation}
f(\theta)=\frac{1}{3}\cos\theta\left(1+2\cos\theta\right).
\label{eq:ftheta}
\end{equation}

We explore the parameter grid listed in Table~\ref{tab:model_parameters}, with $M_{\rm BH}=10^6$--$10^9\,M_\odot$, $\dot{m}=0.1$--$1$ and a fixed radiative efficiency $\eta=0.3$. The SED shown in Fig.~\ref{AGNpectra} is at $\theta=75^\circ$, and for other inclinations we rescale the bolometric luminosity using Eqs.~(\ref{eq:Ltheta}) and (\ref{eq:ftheta}).

\subsection{Dust in the dusty torus}

The dusty torus in AGN is located close to the central engine and is subject to intense radiation fields. With a typical spatial extent of only a few tens of parsecs, direct observational constraints on the dust composition, size distribution, and spatial structure remain limited by angular resolution and line-of-sight mixing. Although recent advances in infrared interferometry and high-resolution imaging have improved our understanding of the torus morphology, detailed information about the physical nature of the dust grains remains elusive.

Given these observational limitations, AGN radiative-transfer calculations commonly adopt simplified dust prescriptions motivated by the diffuse interstellar medium (ISM). A widely used reference is a silicate--carbonaceous grain population with an MRN-type size distribution, $n(a)\propto a^{-3.5}$ \citep{1977ApJ...217..425M}. In this work, we instead use Astrodust as a fixed, diffuse-ISM-motivated opacity prescription. This provides a controlled baseline for isolating the effects of geometry, optical depth, and radial density, but it should not be interpreted as a complete description of dust processed in the AGN environment.

\subsubsection{Dust properties}
\label{sec:dust_properties}
We adopt the \texttt{Astrodust} grain model of \citet{Draine_2021} for the dusty torus. The corresponding optical properties reproduce the main MIR silicate features commonly discussed in AGN torus models \citep[e.g.][]{Fritz2006,Draine2007,Nenkova2008}. We use the \texttt{Astrodust} material density $\rho_{\rm Ad}=2.74~{\rm g\,cm^{-3}}$.

For a grain size distribution discretised in bins $a_i$, we compute the mass-weighted absorption coefficient as
\begin{equation}
\kappa_{\mathrm{abs}}(\lambda) = \sum_i w_{i}^m \,\kappa_{\mathrm{abs}}(a_i,\lambda),
\label{wkappa}
\end{equation}
with
\begin{equation}
w_{i}^m = \frac{m(a_i)\,n(a_i)}{\sum_i m(a_i)\,n(a_i)} ,
\end{equation}
where $m(a_i)$ is the mass of a grain of size $a_i$ and $n(a_i)$ is the corresponding number density.

Since scattering is not treated explicitly, we approximate its impact on attenuation by adopting the effective absorption coefficient \citep{1979rpa..book.....R}
\begin{equation}
\kappa_{\mathrm{eff}}(\lambda) = \sqrt{\kappa_{\mathrm{abs}}(\lambda)\,\kappa_{\mathrm{ext}}(\lambda)} .
\label{kappa_eff}
\end{equation}

\begin{figure}[t]
\centering
\includegraphics[width=\linewidth]{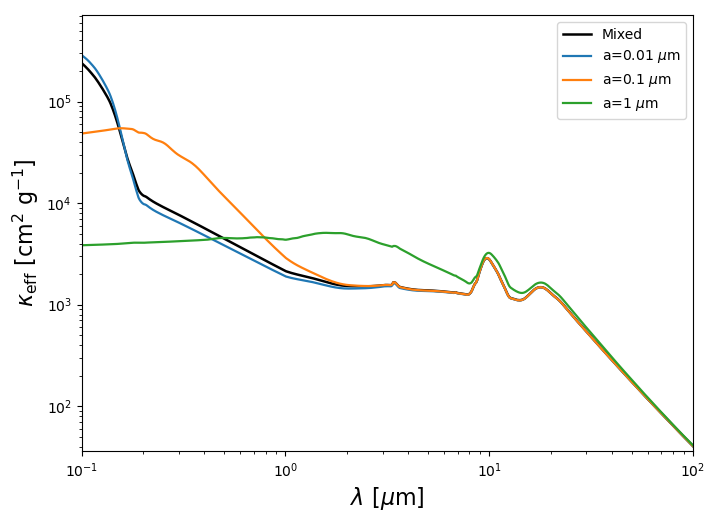}
\caption{Effective absorption coefficient $\kappa_{\mathrm{eff}}$ of \texttt{Astrodust} as a function of wavelength for selected grain sizes and the adopted mixed population \citep{Hensley_2023}.}
\label{kappa}
\end{figure}

Figure~\ref{kappa} shows $\kappa_{\mathrm{eff}}(\lambda)$ for selected grain sizes and for the full size distribution used in this work \citep{Hensley_2023}.

\subsubsection{Dusty torus geometry}
\label{dusty torus geometry}
The geometry of the dusty torus is primarily governed by the physical conditions required for dust survival and by dynamical constraints imposed by the environment. The innermost boundary of the dusty structure is set by the dust sublimation radius, which corresponds to the location where the heating and cooling of dust grains are in radiative equilibrium. Dust grains can survive at this radius provided their temperature does not exceed the sublimation threshold \citep{Dwek1987}.

The radiative heating rate of dust grains at a distance \( D \) from the central source and at an inclination angle \( \theta \) is given by
\begin{equation}
    H_\mathrm{d}(\theta, D) = \int \bar{m} \kappa_{\mathrm{eff}}(\lambda) \frac{L^\mathrm{s}_\lambda(\lambda, \theta)}{4 \pi D^2}  \mathrm{d}\lambda,
\label{Heating}
\end{equation}
where $\bar{m}$ is the average mass and \( L^\mathrm{s}_\lambda(\lambda, \theta) \) is the monochromatic luminosity of the central source as seen from angle \( \theta \). The corresponding cooling rate due to thermal re-emission is
\begin{equation}
    L_\mathrm{d}(T_\mathrm{d}) = \int 4\pi \bar{m} \kappa_{\mathrm{eff}}(\lambda) B_\lambda(\lambda, T_\mathrm{d})  \mathrm{d}\lambda,
\label{Cooling}
\end{equation}
where \( B_\lambda(\lambda, T_\mathrm{d}) \) is the Planck function at the dust temperature \( T_\mathrm{d} \). By equating the heating and cooling terms, the dust sublimation radius for a temperature of \( T_\mathrm{d} = 1500\mathrm{K} \) can be expressed as
\begin{equation}
    R_{1500K}(\theta) = \sqrt{
        \frac{
            \int \bar{m} \kappa_{\mathrm{eff}}(\lambda) L^s_\lambda(\lambda, \theta)  \mathrm{d}\lambda
        }{
            4\pi L_d(1500\mathrm{K})
        }
    }.
\end{equation}

This defines the angle-dependent inner boundary, $R_{\mathrm{in}}(\theta)$, at which grains with the adopted opacity can remain thermally stable at $T_{\rm sub}=1500\,\mathrm{K}$. Material interior to this boundary is treated as dust-free, while dust is permitted at larger radii. The calculation prescribes a sublimation boundary and does not model grain condensation or destruction kinetics.

The spatial extent of the torus is set by its inner and outer radii. The outer radius is parameterised as
\begin{equation}
R_{\mathrm{out}}(\theta) = Y\,R_{\mathrm{in}}(\theta),
\end{equation}
where $Y$ is a dimensionless radial extent parameter (Table~\ref{tab:model_parameters}). We assume an axisymmetric torus with an opening angle $\theta_{\mathrm{open}}$; the polar cone ($\theta < \theta_{\mathrm{open}}$) is taken to be dust-free.

\subsubsection{Dust spatial distribution and density laws}
\label{sec:density_structure}

The torus emission depends sensitively on how dust is distributed in space. In wind-driven scenarios, dust is expected to be embedded in the outflowing gas and approximately co-spatial with it over the region where grains can survive. We therefore prescribe the gas density structure and convert it to a dust density via a constant dust-to-gas mass ratio, $\mathcal{D}$.

We adopt a parameterised density law motivated by the self-similar MHD disk-wind solution of \citet{Fukumura2010}. We assume axisymmetry and define the dimensionless radius $X\equiv r/R_{\rm sch}$, where $R_{\rm sch}$ is the Schwarzschild radius of the central black hole. In spherical coordinates $(r,\theta)$, the hydrogen nuclei number density is written as
\begin{equation}
    n(r,\theta) = \mathcal{N}(p,\theta)\, n_0 \,\frac{\dot{m}}{M_8}\, X^{-p}
    \exp\!\left[\frac{5}{2}\left(\theta - \frac{\pi}{2}\right)\right],
\label{eq:densitylaw}
\end{equation}
where $n_0 = 5 \times 10^{10}~\mathrm{cm}^{-3}$ is a normalisation constant, $M_8 \equiv M_{\rm BH}/(10^8\,M_\odot)$. The parameter $p$ controls the radial density gradient. We adopt $p=1$ as a fiducial value, consistent with a commonly used self-similar MHD-wind scaling \citep{1982MNRAS.199..883B,Sarangi_2019}. For comparison, $p=2$ has the same radial scaling as a steady, constant-velocity spherical outflow \citep{2000ApJ...545...63E}. We additionally explore ($p=0.5$ and $p=1.5$) to span a broader range of gradients commonly adopted in dusty torus models \citep[e.g.,][]{Fritz2006,Stalevski_2012,2005ApJ...631..689E}.

Changing $p$ at fixed normalisation modifies not only the radial distribution but also the total column intercepted along a given direction. To isolate the effect of the density gradient, we enforce that the radial hydrogen column density
\begin{equation}
    N_{\rm H}(\theta) = \int_{R_{\mathrm{in}}(\theta)}^{R_{\mathrm{out}}(\theta)} n(r, \theta)\, \mathrm{d}r
\end{equation}
is conserved at each polar angle $\theta$ relative to the fiducial $p=1$ model. Defining $X_{\rm in}(\theta)\equiv R_{\rm in}(\theta)/R_{\rm Sch}$, this requirement sets the $p$- and angle-dependent normalisation factor $\mathcal{N}(p,\theta)$ to
\begin{equation}
    \mathcal{N}(p,\theta) = X_{\rm in}^{p-1}(\theta)\,\frac{(1-p)\ln(Y)}{Y^{1-p}-1}, \qquad p\neq1,
\end{equation}
with $\mathcal{N}(1,\theta)=1$. With this choice, models with different $p$ share the same $N_{\rm H}(\theta)$ (for fixed $Y$) while differing only in how that column is distributed with radius.

In the shielded dusty regions relevant for the torus, the gas mass is dominated by hydrogen; we therefore identify $n(r,\theta)\equiv n_{\rm H}(r,\theta)$. The dust mass density is then
\begin{equation}
    \rho_{\mathrm{dust}}(r,\theta) = \mathcal{D}\, m_{\mathrm{H}}\, n(r,\theta),
\end{equation}
where $m_{\rm H}$ is the mass of a hydrogen atom. We explore $\mathcal{D}$ in the range $10^{-4}$--$10^{-2}$, spanning dust-poor conditions to values comparable to the diffuse Milky Way ISM \citep[e.g.,][]{Draine2007,Remy-Ruyer2014,Valiante2011}. The full set of model parameters is summarised in Table~\ref{tab:model_parameters}.

\begin{table}[t]
\caption{Model parameters}
\label{tab:model_parameters}
\centering
\begin{tabular}{lp{0.63\columnwidth}}
\hline\hline
Parameter & Adopted values \\
\hline
$M_{\rm BH}$ & $10^6$, $5\times 10^6$, $10^7$, $5\times10^7$, $10^8$, $5\times10^8$, $10^9\,M_\odot$ \\
$\dot{m}$ & $0.1,\,0.3,\,0.5,\,0.8,\,1.0$ \\
$\log\mathcal{D}$& -2.0, -2.4,  -2.8, -3.2,  -3.6, -4.0\\
$\eta$ & 0.3\\
$\alpha_{\rm OX}$ & -1.5 \\
$\theta_{\mathrm{open}}$& $20^\circ$, $30^\circ$, $40^\circ$, $50^\circ$, $60^\circ$\\
$\theta_{\mathrm{view}}$& $0^\circ$, $10^\circ$, $20^\circ$, $30^\circ$, $40^\circ$, $50^\circ$, $60^\circ$, $70^\circ$, $80^\circ$, $90^\circ$\\
$R_{\mathrm{in}}$& $R_{1500\mathrm{K}}$\\
$Y$ &20, 60, 100,140,180\\
\hline
\end{tabular}
\end{table}

\subsubsection{Radiative-transfer modelling}
\label{sec:numerical}

We compute the dust temperature distribution and emergent SEDs with a deterministic 3D ray-tracing scheme on an axisymmetric dusty torus. The dust optical properties are taken from the Astrodust grain model (Sect.~\ref{sec:dust_properties}) Throughout the radiative transfer we use a wavelength-dependent effective opacity, $\kappa_{\rm eff}(\lambda)$, evaluated with the same mass-weighted prescription as Eq.~(\ref{wkappa}) but using the effective cross sections adopted in our scheme. For a dust mass column density $\Sigma_{\rm d}$, the optical depth is
\begin{equation}
\tau_\lambda=\kappa_{\rm eff}(\lambda)\,\Sigma_{\rm d}.
\end{equation}

At each location, the dust temperature $T_{\rm d}$ is obtained by imposing local radiative equilibrium. Using the heating and cooling rates (Eqs.~\ref{Heating} and \ref{Cooling}), we solve for $T_{\rm d}$ such that the absorbed and emitted powers balance. The calculation includes direct heating by the primary AGN radiation but does not iteratively include the diffuse infrared radiation field produced by other dust cells (dust self-heating). This approximation is most relevant for optically thick, centrally concentrated models, in which reprocessed radiation can heat shielded regions and modify the MIR SED. The results should therefore be interpreted as a controlled comparison of radial density laws within the adopted direct-heating approximation. A fully iterative treatment of the temperature field is left for future work.

Synthetic SEDs are computed by a formal solution along rays toward an observer at viewing angle $\theta_{\rm view}$. The total observed luminosity is obtained by summing the contribution of all volume elements, where each cell emits thermally and is attenuated by the optical depth accumulated from the cell to the outer boundary along the line of sight. Denoting the dust mass density by $\rho_{\rm dust}$, the cell emissivity is taken to be proportional to $\rho_{\rm dust}\,\kappa_{\rm eff}(\lambda)\,B_\lambda(\lambda,T_{\rm d})$. The emitted luminosity can be written as
\begin{align}
L_{\lambda}^{\rm obs}(\theta_{\rm view})
&= \int_V 4\pi\,\rho_{\rm dust}(r,\theta)\,\kappa_{\rm eff}(\lambda)\,
B_\lambda(\lambda,T_{\rm d})
\nonumber\\
&\qquad \times \exp\!\left[-\tau_\lambda(r,\theta,\theta_{\rm view})\right]\,{\rm d}V .
\label{eq:formal_solution}
\end{align}

The computational domain is discretised on a spherical polar grid $(r,\theta,\phi)$ with logarithmic spacing in $r$ to resolve the strong gradients near the sublimation front. In the calculations presented here we use $N_r=100$, $N_\theta=100$, and $N_\phi=20$.

As a consistency check, we verify global energy conservation by confirming that the total absorbed primary luminosity is balanced by the integrated dust infrared emission.

\subsection{Starburst}
Table~\ref{tab:sb_refs_min} lists the J2000 coordinates and systemic redshifts (z) of the galaxies used as starburst (SB) templates in this work. With the exception of Arp\,220, M\,82, and NGC\,253, all objects are selected from the Spitzer/IRS starburst sample of \citet{2006ApJ...653.1129B} and are classified there as starburst-dominated systems. In our analysis, these templates provide only the SED shapes; the absolute normalisation is determined during the fit to each target.

\begin{figure}[t]
\centering
\includegraphics[width=\linewidth]{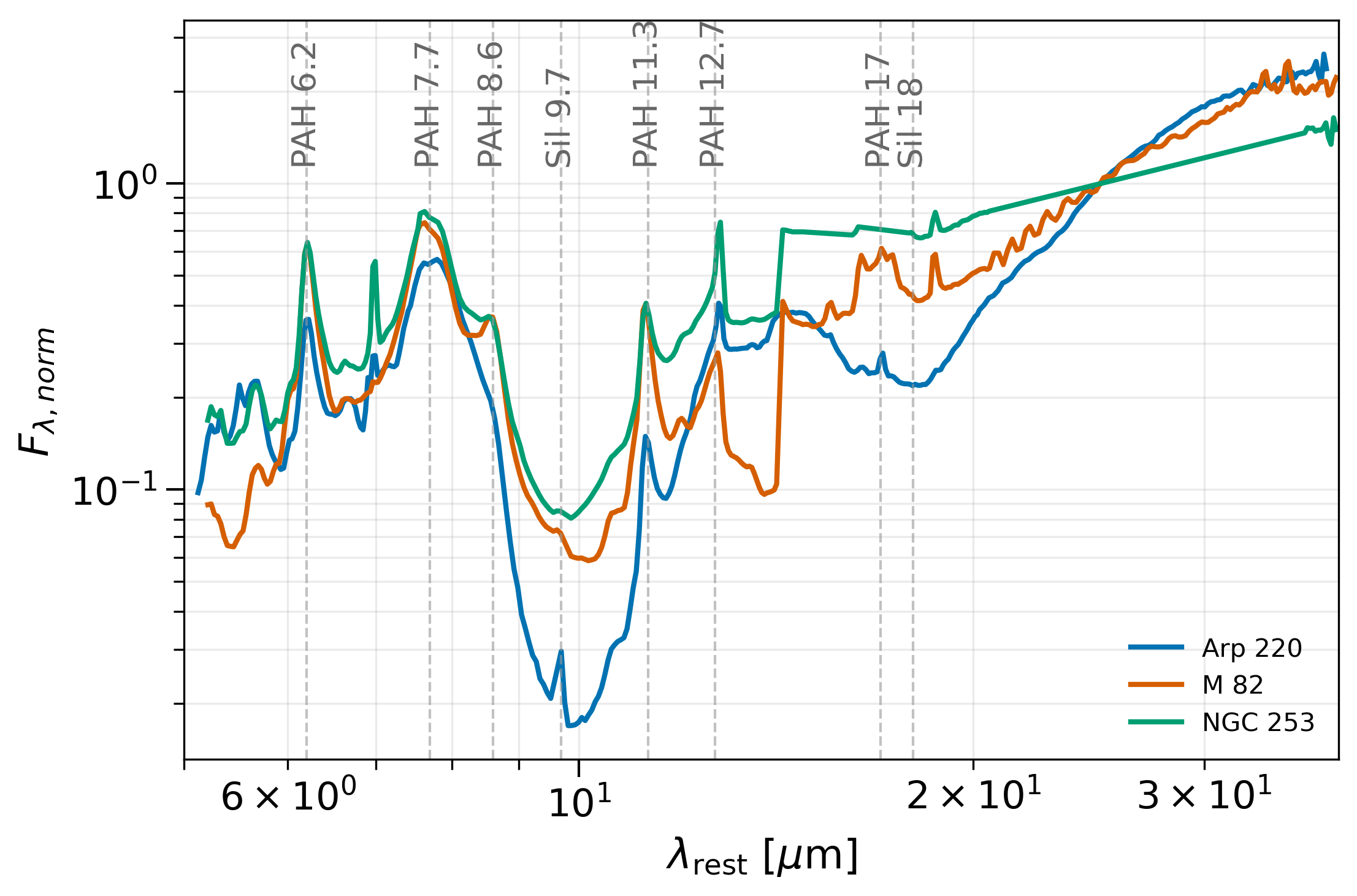}
\caption{Rest-frame MIR starburst templates used in the spectral decomposition. The normalised spectra of Arp~220, M~82, and NGC~253 span different PAH strengths, silicate absorption depths, and continuum shapes.}
\label{fig:sb_templates}
\end{figure}

In addition, we include three well-studied systems, Arp\,220, M\,82, and NGC\,253, as representative SB templates. Arp\,220 is an ultraluminous infrared galaxy (ULIRG), $L_{\mathrm{IR}}\gtrsim10^{12}\,L_{\odot}$, from a late-stage merger with extreme dust obscuration and a very compact nuclear starburst; its MIR shows deep 9.7\,$\mu$m silicate absorption and weak apparent PAH features, and its SED peaks sharply in the far-IR \citep{SandersMirabel1996ARAandA34,DownesSolomon1998ApJ507615,2007ApJ...663...81P}. M\,82 is the nearby prototypical starburst, with strong PAH emission, prominent MIR fine-structure lines, moderate silicate absorption, and a relatively warm far-IR colour; we follow common empirical representations \citep{1998ApJ...509..103S,ForsterSchreiber2003AA399833,2007ApJ...663...81P}. NGC\,253 is a nearby dusty nuclear starburst with strong PAH features and bright far-IR/submillimeter emission; its Spitzer/IRS spectrum places it firmly among starburst-dominated systems \citep{2006ApJ...653.1129B,2001A&A...377...73R,Dale2012ApJ74595}. Together these three templates span a useful range of obscuration, PAH strength, and dust temperature, and thus provide complementary SED shapes for our fits (Fig.~\ref{fig:sb_templates}).

\begin{table}[t]
\centering
\caption{Reference starburst galaxies}
\label{tab:sb_refs_min}
\small
\setlength{\tabcolsep}{6pt}
\begin{tabular}{l l l S[table-format=1.5]}
\toprule
Object & RA (J2000) & Dec (J2000) & {$z$} \\
\midrule
Arp\,220   & 15:34:57.2 & +23:30:11 & 0.0181 \\
M\,82      & 09:55:52.7 & +69:40:46 & 0.0007 \\
Mrk\,52    & 12:25:42.7 & +00:34:20 & 0.0079 \\
NGC\,253   & 00:47:33.1 & $-$25:17:18 & 0.0008 \\
NGC\,1614  & 04:33:59.8 & $-$08:34:44 & 0.0159 \\
NGC\,7714  & 23:36:14.1 & +02:09:19 & 0.0093 \\
IRAS\,17208$-$0014&17:23:21.9 & $-$00:17:00&0.0428\\
\bottomrule
\end{tabular}
\tablefoot{Values compiled from commonly used NED/SIMBAD entries and rounded.}
\end{table}

Figure~\ref{fig:sb_templates} compares these three representative SB templates after uniform preprocessing and normalisation. All spectra have been shifted to the rest frame and resampled onto our fitting wavelength grid. For visualisation, each template is normalised by its rest-frame flux density at $25\,\mu$m,
$F_{\mathrm{norm}}(\lambda)=F_{\lambda}(\lambda)/F_{\lambda}(25\,\mu\mathrm{m})$, so that the curves intersect at unity at $25\,\mu$m; the absolute scaling is free in the fits. Vertical dashed lines mark the principal PAH and silicate features.

In the following section we describe how these SB templates are combined with the AGN torus templates and fitted to the MIRI/MRS spectra.

\subsection{Spectral decomposition and grid-based inference}
\label{sec:decomposition}

To model the rest-frame $5$--$14\,\mu\mathrm{m}$ spectra, we add an AGN torus template and an empirical host-galaxy SB template with non-negative coefficients:
\begin{equation}
\lambda F^{\mathrm{mod}}_{\lambda}=c_{\mathrm{AGN}}\,\lambda F_{\lambda}^{\mathrm{AGN}}
+c_{\mathrm{SB}}\,\lambda F_{\lambda}^{\mathrm{SB}},
\label{eq:decomposition}
\end{equation}
where $c_{\rm AGN},c_{\rm SB}\geq0$ set the strengths of the two components.

We calculate the AGN and SB fractions from their integrated fluxes over $5$--$14\,\mu\mathrm{m}$:
\begin{equation}
f_i=\frac{\displaystyle\int_{5\,\mu\mathrm{m}}^{14\,\mu\mathrm{m}} c_i\,\lambda F_{\lambda,i}(\lambda)\,\mathrm{d}\lambda}{\displaystyle\sum_j\int_{5\,\mu\mathrm{m}}^{14\,\mu\mathrm{m}} c_j\,\lambda F_{\lambda,j}(\lambda)\,\mathrm{d}\lambda}.
\end{equation}

\subsubsection{Likelihood}
\label{sec:norm_nnls_like}

Before fitting, we divide the observed JWST spectrum and each template by its median flux over the fitted wavelength range. For every AGN+SB template pair, we determine $(c_{\rm AGN},c_{\rm SB})$ using non-negative least squares (NNLS).

The likelihood is taken to be Gaussian, with an effective uncertainty that combines the statistical error with a 10\% fractional term:
\begin{equation}
\sigma_{\lambda,\mathrm{eff}}^{2} = \sigma_{\lambda,\mathrm{stat}}^{2} + \left(f_{\mathrm{sys}}\,\lambda F^{\mathrm{obs}}_{\lambda}\right)^{2},
\end{equation}
where $f_{\mathrm{sys}}=0.1$. The log-likelihood is
\begin{equation}
\ln \mathcal{L} = -\frac{1}{2}\sum_{\lambda}\left[\frac{\left(\lambda F^{\mathrm{obs}}_{\lambda}-\lambda F^{\mathrm{mod}}_{\lambda}\right)^{2}}{\sigma_{\lambda,\mathrm{eff}}^{2}} + \ln\left(2\pi\sigma_{\lambda,\mathrm{eff}}^{2}\right)\right].
\end{equation}

\subsubsection{Priors and grid weights}

For each evaluated grid point, we combine the likelihood and prior as
\begin{equation}
\ln P(\Theta \mid D) =
\ln \mathcal{L}(D \mid \Theta) + \ln \pi(\Theta) + \mathrm{const.}
\end{equation}
This ordinary posterior corresponds to $\beta=1$ and is used for the fiducial results. In the sensitivity tests, we instead use powered grid weights,
\begin{equation}
w_{\beta,k}\propto \mathcal{L}_k^\beta \pi_k ,
\label{eq:beta_weight}
\end{equation}
where $k$ labels a grid point and $\beta$ controls the relative weight given to the spectral likelihood compared with the prior. The parameter set $\Theta$ contains the AGN grid parameters and the SB template identity; the two component coefficients are determined by NNLS. We adopt flat discrete priors over the sampled values of $M_{\rm BH}$, $\theta_{\rm view}$, $\theta_{\rm open}$, $Y$, and $\mathcal{D}$, and over the seven SB templates.

For $\dot m$, we use the intrinsic all-AGN Eddington-ratio distribution function (ERDF) of \citet{2022ApJS..261....9A}:
\begin{equation}
\xi(\log\lambda_{\rm Edd})\equiv
\frac{\mathrm{d}N}{\mathrm{d}\log\lambda_{\rm Edd}}
\propto
\left[
\left(\frac{\lambda_{\rm Edd}}{\lambda_*}\right)^{\delta_1}
+
\left(\frac{\lambda_{\rm Edd}}{\lambda_*}\right)^{\delta_2}
\right]^{-1},
\end{equation}
with $\log\lambda_*=-1.338$, $\delta_1=0.38$, and $\delta_2=2.64$. We relate the Eddington ratio to the model accretion rate through $\lambda_{\rm Edd}=\eta\dot m$, with $\eta=0.3$. Because the ERDF is defined per logarithmic interval, the prior weight of each $\dot m_i$ value is proportional to $\xi(\log\lambda_i)\Delta\log\lambda_i$, where $\lambda_i=\eta\dot m_i$. These weights are normalised over the five available $\dot m$ values. We impose no additional $M_{\rm BH}$-dependent limits on $\lambda_{\rm Edd}$ or $\dot m$.

For each radial density law, we calculate the grid score and normalised weight as
\begin{equation}
\mathcal{Z}(p)=\sum_{k\in p}\mathcal{L}_k\pi_k,
\qquad
W(p\mid D)=
\frac{\mathcal{Z}(p)}
{\sum_{p'}\mathcal{Z}(p')},
\label{eq:discrete_evidence}
\end{equation}
where $\mathcal{L}_k$ is the likelihood after fitting the component coefficients and $\pi_k$ is the prior probability of grid point $k$. Because the model library is finite and precomputed, the score is obtained by direct summation over all evaluated grid points. This differs from methods such as Markov chain Monte Carlo and nested sampling, which draw samples from a parameter space and are mainly useful when the model parameters are treated as continuous variables \citep[e.g.][]{MacKay2003,Gregory2005,Skilling2004,Trotta2008}.

We use $\beta=1$ for the main results and repeat the calculation with $\beta=0.001$, 0.003, 0.01, 0.03, and 0.1 as sensitivity tests. Table~\ref{tab:density_laws_by_object_bayes} reports the weighted median parameters for each density law. The listed $\chi^2$ value and SB template correspond to the maximum-a-posteriori (MAP) grid point.

\section{Sample and observational data}
\label{sec:sample}

We assembled a working sample of 25 AGN with publicly available JWST/MIRI-MRS observations and the ancillary information required for the model priors. The sample overlaps with the nearby AGN and type~2 quasar samples discussed by \citet{2025MNRAS.539.2158G} and \citet{2025AA...698A.194R,2025arXiv251202629R}. The spectra used here were retrieved directly from the Mikulski Archive for Space Telescopes (MAST) as standard JWST pipeline one-dimensional products.

\subsection{Sample selection and archive spectra}

The objects listed in Table~\ref{tab:jwst_sample_extended} span a broad range of obscuration levels and host-galaxy environments, including nearby Seyfert galaxies, low-luminosity AGN, luminous infrared galaxies, and five optically selected type~2 quasars from QSOFEED. Black hole mass estimates from the literature span approximately $10^5$--$10^9\,M_\odot$..

All targets were observed with JWST/MIRI MRS. We use the calibrated \texttt{x1d} products from MAST, corresponding to JWST pipeline \texttt{EXTRACT1D} spectra with data model \texttt{MRSMultiSpecModel} \citep{Bushouse2024}. The products are marked in their archive headers as \texttt{SRCTYPE=EXTENDED} and were generated with pipeline version 1.20.2 and Calibration Reference Data System (CRDS) context \texttt{jwst\_1464.pmap}. The observations come from JWST programmes 1039, 1268, 1328, 1670, 1717, 1875, 2004, 2016, 2732, 2773, 3655, 4225, and 6138.

For products classified as extended sources, \texttt{EXTRACT1D} sums the reconstructed integral-field-unit image and performs no local background subtraction. We did not re-extract the data cubes with a custom aperture or apply a point-spread-function (PSF) decomposition such as MRSPSFisol \citep{2025MNRAS.539.2158G}. Because both the MRS field of view and the PSF vary with channel and wavelength, these spectra are neither matched-aperture nor PSF-isolated measurements of the unresolved nucleus. Host-galaxy and circumnuclear emission may therefore remain. The empirical SB component represents part of this contribution spectrally, but it cannot correct for all effects of the wavelength-dependent extraction area.

We applied the same post-processing to all archive spectra. Each of the 12 MRS sub-bands was corrected for Galactic foreground extinction using the listed $E(B-V)$ values, $A_V=3.1E(B-V)$, and the Milky Way average MIR extinction curve of \citet{2021ApJ...916...33G}, shifted to the rest frame using the redshifts in Table~\ref{tab:jwst_sample_extended}, converted to $\lambda F_{\lambda}$, and interpolated onto a uniform $0.05\,\mu\mathrm{m}$ grid. Measurements in overlapping sub-bands were combined with inverse-variance weights, with the weighted sub-band scatter added in quadrature to the statistical uncertainty. The empirical \textit{Spitzer} starburst templates were shifted to the rest frame and resampled onto the same grid before fitting (\citealt{2006ApJ...653.1129B}; Table~\ref{tab:sb_refs_min}). Before fitting, we removed small wavelength intervals around the main narrow MIR emission lines, while retaining the broad PAH complexes so that they could be represented by the SB templates.

The fiducial interval is the rest-frame $5$--$14\,\mu\mathrm{m}$ range, which includes the short-wavelength continuum and the diagnostic $9.7\,\mu\mathrm{m}$ silicate feature. Trial fits extending to $20\,\mu\mathrm{m}$ require an additional contribution from cool dust beyond approximately $14\,\mu\mathrm{m}$. Because the dust mass is coupled to the accretion-linked gas density, $\mathcal{D}$, and the adopted geometry, increasing $\mathcal{D}$ alone would also change the silicate optical depth. We therefore use $5$--$14\,\mu\mathrm{m}$ for the quantitative model comparison and leave a more flexible treatment of the cool-dust emission to future work.

\section{Results}
\label{3}
\subsection{Spatial and temperature distribution}
\begin{figure}[t]
    \centering
    \includegraphics[width=\linewidth]{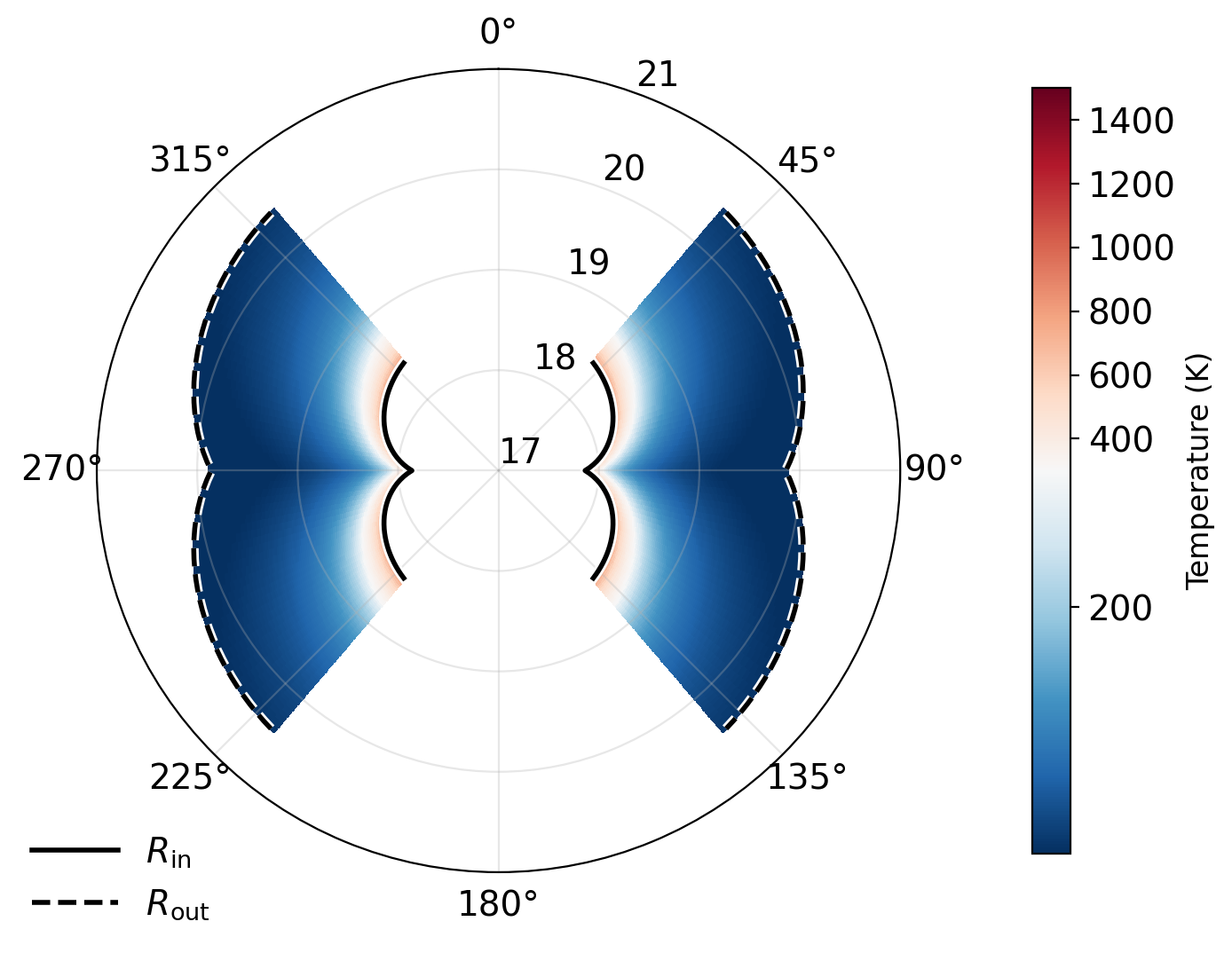}
    \caption{Geometry--temperature polar map of the dusty torus for the representative model with $M_{\rm BH}=10^{8}\,M_\odot$, $\dot m=0.5$, $\theta_{\rm open}=40^\circ$, $Y=100$, and ${\mathcal{D}}=10^{-2.8}$. Colours indicate the dust temperature and the radial coordinate is shown as $\log_{10}(r/{\rm cm})$. At each polar angle $\theta$, the plotted domain is restricted to the dusty region between the sublimation front $R_{\rm in}(\theta)$ (black solid curve; defined by $T_{\rm dust}=1500\,{\rm K}$) and the outer boundary $R_{\rm out}(\theta)$ (black dashed curve), with $R_{\rm out}(\theta)=Y\,R_{\rm in}(\theta)$ by construction. The blank angular sectors correspond to the adopted polar opening cones where dust is excluded.}
    \label{T}
\end{figure}

Figure~\ref{T} shows the dust temperature field \(T(r,\theta)\) for a representative model (\(M_{\rm BH}=10^{8}\,M_\odot\), \(\dot m=0.5\), \(\theta_{\rm open}=40^\circ\), \(Y=100\), \(\mathcal{D}=10^{-2.8}\)), plotted between the sublimation surface \(R_{\rm in}(\theta)\) and \(R_{\rm out}(\theta)=Y\,R_{\rm in}(\theta)\).

The temperature field exhibits strong angular anisotropy, with more efficient heating at higher latitudes than near the equatorial plane. This anisotropic illumination imprints directly on the torus inner rim: the sublimation front $R_{\rm in}(\theta)$ becomes distinctly non-spherical and assumes a concave, bowl-like shape, with larger radii towards polar directions than along the equator.

Across the model grid, geometric and opacity parameters primarily control the temperature extent and the relative weight of hot versus cool dust. Increasing \(Y\) extends the cool outer region, while larger \(\mathcal{D}\) increases attenuation and steepens the radial temperature decline. These trends are reflected in the SED variations discussed below (Fig.~\ref{fig:parameter_study}).

\subsection{SED library: overview and parameter sensitivity}
\label{sec:model_lib}

\begin{figure*}[t]
\centering
\includegraphics[width=\linewidth]{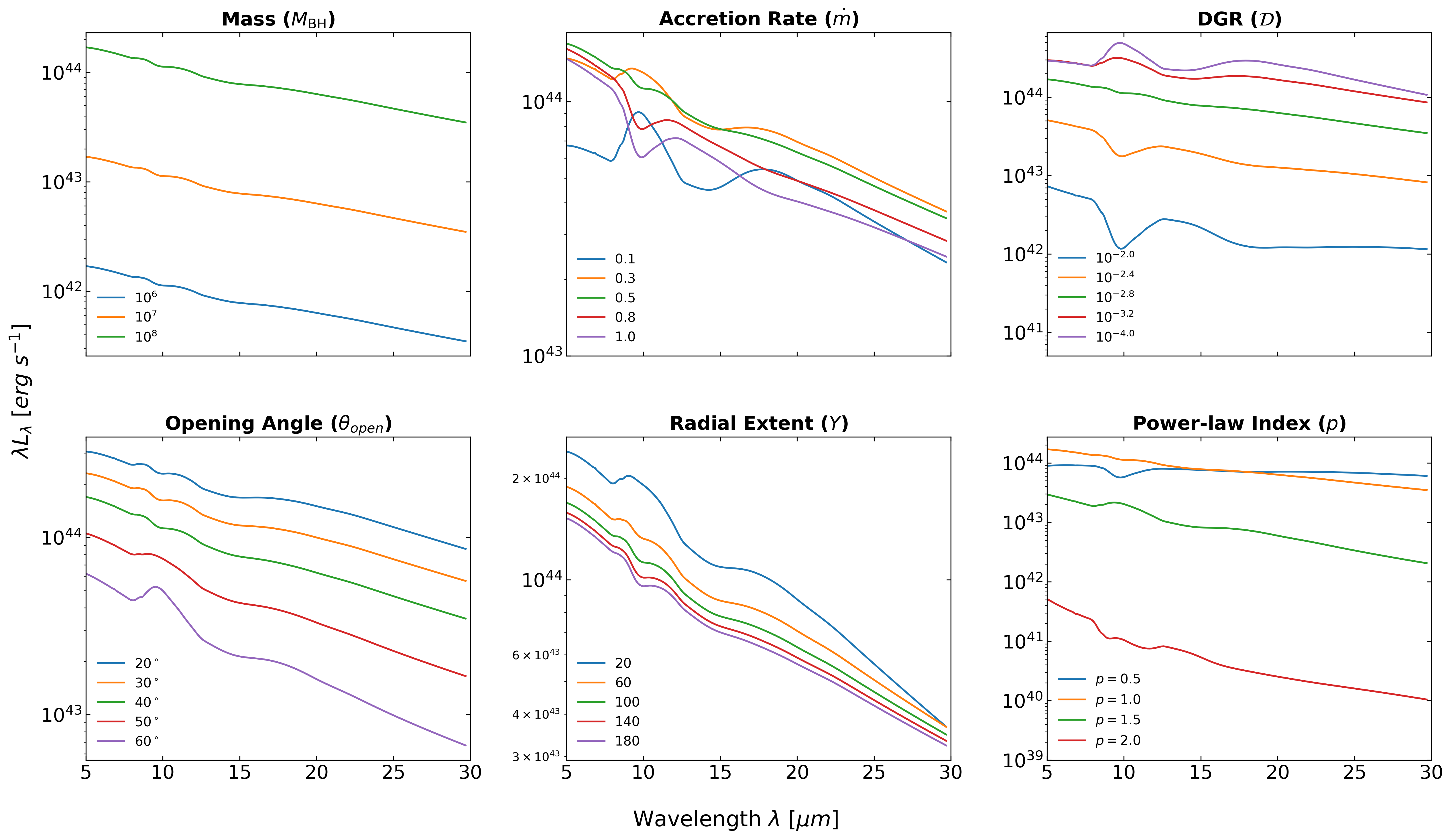}
\caption{One-parameter sensitivity of the infrared SED around the fiducial model. Each panel varies one physical parameter while holding the others fixed, illustrating changes in both the continuum and the $9.7\,\mu\mathrm{m}$ silicate feature.}
\label{fig:parameter_study}
\end{figure*}

We build a physically motivated library of AGN SEDs over a wide parameter space. The grid spans seven black hole masses, $M_{\rm BH}\in[10^{6},10^{9}]\,M_\odot$. It also varies the normalised accretion rate $\dot{m}$, the dust-to-gas ratio $\mathcal{D}$, the torus opening angle $\theta_{\rm open}$, the radial extent $Y\equiv R_{\rm out}(\theta)/R_{\rm in}(\theta)$, the viewing angle $\theta_{\rm view}$, and the radial density index $p$. This library is used as the forward model in Sect.~\ref{sec:decomposition}.

Figure~\ref{fig:grid_spectra} shows the MIR behaviour of the library for a representative subset. Across the library, the spectra are anisotropic at fixed intrinsic parameters. The 9.7~$\mu$m silicate feature changes with viewing angle. It can appear in emission for more face-on lines of sight and in absorption for more edge-on lines of sight. The orientation contrast becomes stronger when $\mathcal{D}$ is higher. In that case, obscured views show deeper silicate absorption.

We then test parameter sensitivity around a fiducial model (Fig.~\ref{fig:parameter_study}). We vary one parameter at a time and keep the others fixed. Changing $M_{\rm BH}$ mainly shifts the overall normalisation. The MIR spectral shape changes only weakly. Changing $\dot m$ affects the continuum slope and the silicate feature. The trend with $\dot m$ can be non-monotonic in our grid.

Parameters that set optical depth and geometry have the largest impact on the spectral morphology. Increasing $\mathcal{D}$ strengthens silicate self-absorption. Increasing $\theta_{\rm view}$ increases the effective line-of-sight extinction. Changing $\theta_{\rm open}$ modifies the dust covering factor and the relative weight of hot inner emission. Increasing $Y$ adds more cool dust at large radii. This typically reddens the MIR continuum and increases absorption.

Finally, the radial density index $p$ is a key structural parameter. In Fig.~\ref{fig:parameter_study}, increasing $p$ from 0.5 to 2.0 strongly suppresses the MIR output with all other parameters held fixed, and it also changes the continuum slope. In the fits, the component amplitudes are free, so $p$ is constrained mainly by the continuum shape and silicate-profile structure. We therefore treat \(p\) as a structural degree of freedom and assess its impact on the inferred parameters.

\subsection{Spectral decomposition results}
\label{subsec:results_decomp}

\subsubsection{Representative fits and component behaviour}
\label{subsubsec:rep_fits}

\begin{figure*}[t]
    \centering
    \includegraphics[width=\linewidth]{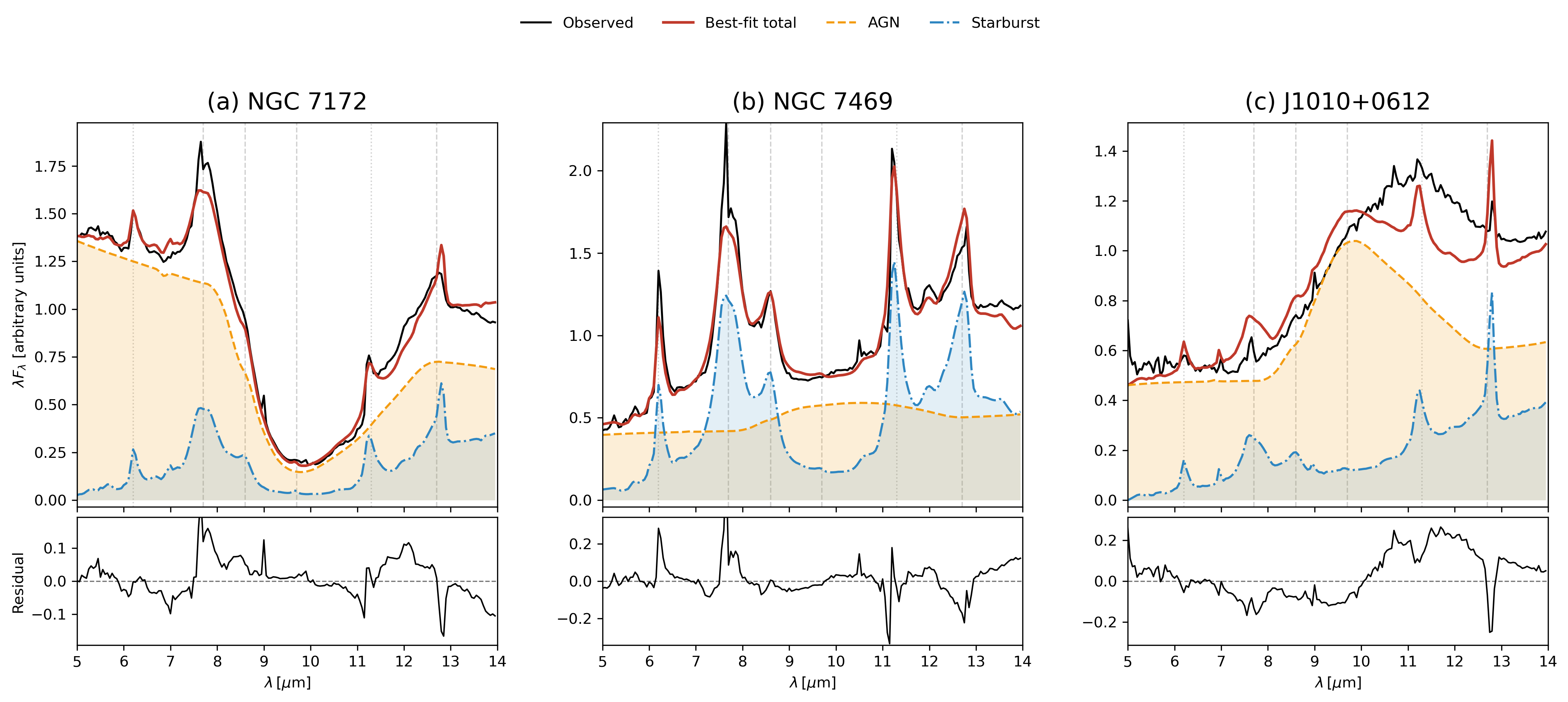}
    \caption{MIR spectral decompositions for (a) NGC~7172, (b) NGC~7469, and (c) J1010+0612. The black and red solid curves show the observed spectrum and fiducial ($\beta=1$) joint MAP model, respectively. Orange dashed and blue dash-dotted curves show the AGN and SB components. The lower panels display the residuals.}
    \label{fig:threeexp}
\end{figure*}

Figure~\ref{fig:threeexp} illustrates the $5$--$14\,\mu\mathrm{m}$ AGN+SB decompositions for three representative sources. The model reproduces the broad continuum and principal PAH and silicate structures.

NGC~7172 represents a heavily obscured source with broad silicate absorption. NGC~7469 and J1010+0612 illustrate cases in which the SB template contributes appreciably to PAH-rich structure while the AGN component sets much of the underlying MIR continuum. The fits are driven by the short-wavelength continuum, PAH complexes, and the main $9.7\,\mu\mathrm{m}$ silicate feature.

\subsection{Global parameter trends and fitting performance}
\label{subsec:global_trends}

\begin{figure*}[t]
    \centering
    \includegraphics[width=\linewidth]{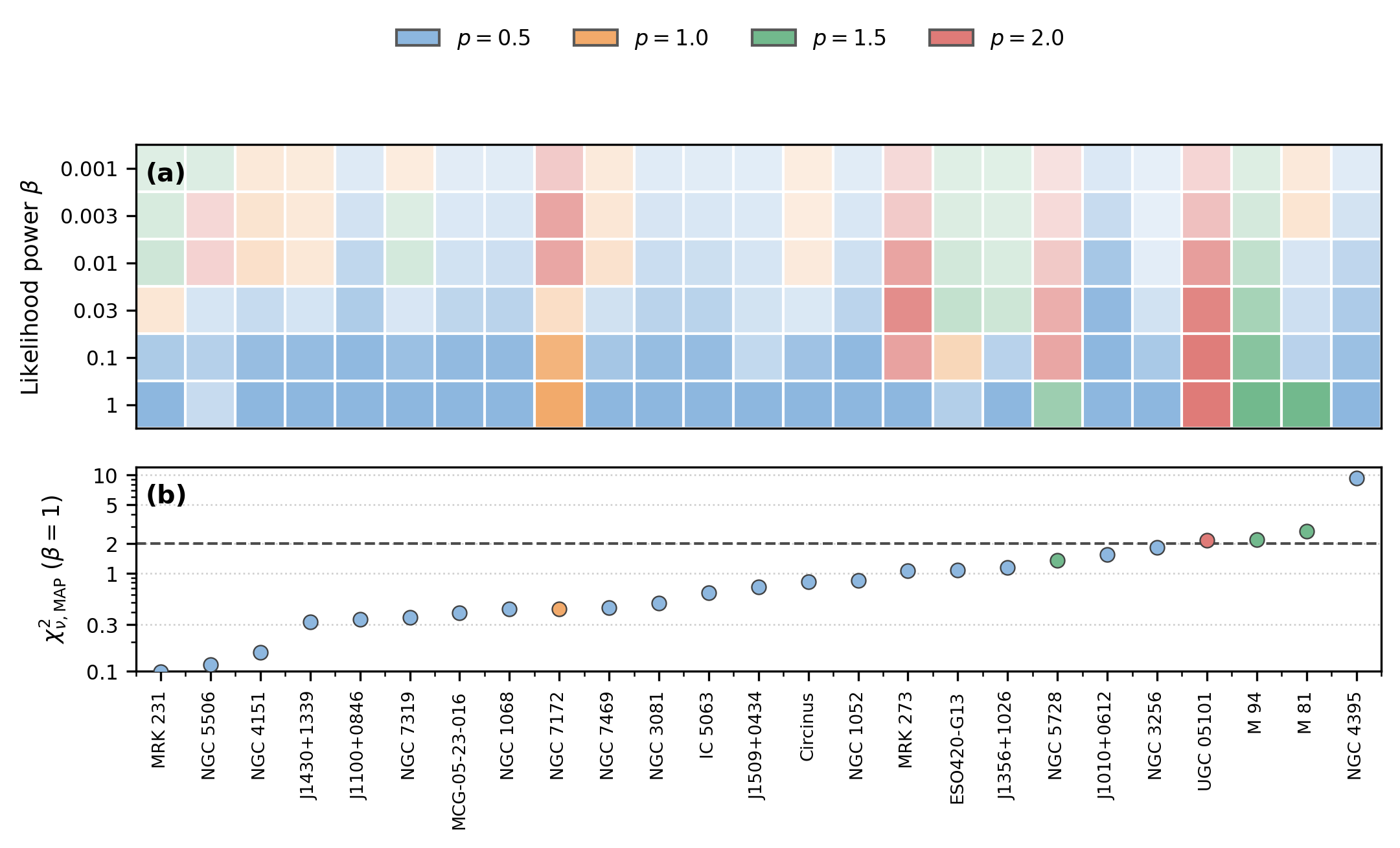}
    \caption{Radial-density ranking and fiducial fit quality. Panel~(a) shows the preferred density law for each source under the ERDF prior for different likelihood powers. Colour identifies the preferred $p$, and colour saturation indicates its normalised grid weight. Panel~(b) shows $\chi^2_{\nu,\mathrm{MAP}}$ for the fiducial $\beta=1$ calculation; the dashed line marks $\chi^2_\nu=2$. Sources are ordered by increasing fiducial reduced chi-squared.}
    \label{globelP}
\end{figure*}

Table~\ref{tab:density_laws_by_object_bayes} lists the grid scores, fit statistics, and parameter summaries for each source and density law. Figure~\ref{globelP} summarises the corresponding density-law rankings and the fiducial fit quality.

For the fiducial $\beta=1$ calculation, 20 of the 25 sources have their largest density-law weight at $p=0.5$. NGC~7172 selects $p=1.0$; M~81, M~94, and NGC~5728 select $p=1.5$; and UGC~05101 selects $p=2.0$. When the two shallow families are combined, 21 sources have $W(p\leq1\mid D)>0.5$. The sample therefore favours shallow radial density laws within the sampled grid.

The MAP fits have a median reduced chi-squared of 0.72. Four sources have $\chi^2_{\nu,\mathrm{MAP}}>2$: UGC~05101, M~94, M~81, and NGC~4395. Among the remaining 21 sources, 19 still select $p=0.5$, while NGC~7172 and NGC~5728 select $p=1.0$ and $p=1.5$, respectively. Thus, the preference for shallow profiles is not driven only by the least well-fitted objects.

The preferred density law of individual sources can change with $\beta$, but the sample-level trend remains. Twelve sources retain the same preferred law across all tested values. The number of sources with $W(p\leq1\mid D)>0.5$ is 17, 17, 17, 20, 21, and 21 for $\beta=0.001$, 0.003, 0.01, 0.03, 0.1, and 1, respectively.

The inferred component balance also depends on $p$. The median $f_{\rm AGN}$ decreases from 0.69 at $p=0.5$ to 0.53, 0.46, and 0.41 at $p=1.0$, 1.5, and 2.0, respectively. Steeper torus profiles therefore require a larger SB contribution on average.

The non-radial grid parameters are less uniquely determined. For the preferred density law of each source, 12 MAP solutions lie at the minimum opening angle, 16 at the maximum radial extent, and 11 at the edge-on viewing angle. These concentrations should be interpreted as model-grid preferences rather than direct measurements of a common obscuring geometry.

Finally, we tested whether the density-law ranking depends on the adopted $\dot m$ prior. Replacing the ERDF prior with equal probabilities for the five discrete $\dot m$ values changes the preferred density law only for NGC~5728, from $p=1.5$ to $p=2.0$. The number of sources with $W(p\leq1\mid D)>0.5$ remains 21, indicating that the main density-law result is not driven by the accretion-rate prior.

\section{Discussion and conclusions}
\label{discussion}

\subsection{The silicate feature: emission vs. absorption}
\label{subsec:silicate_comparison}

\begin{figure*}[t]
    \centering
    \includegraphics[width=\linewidth]{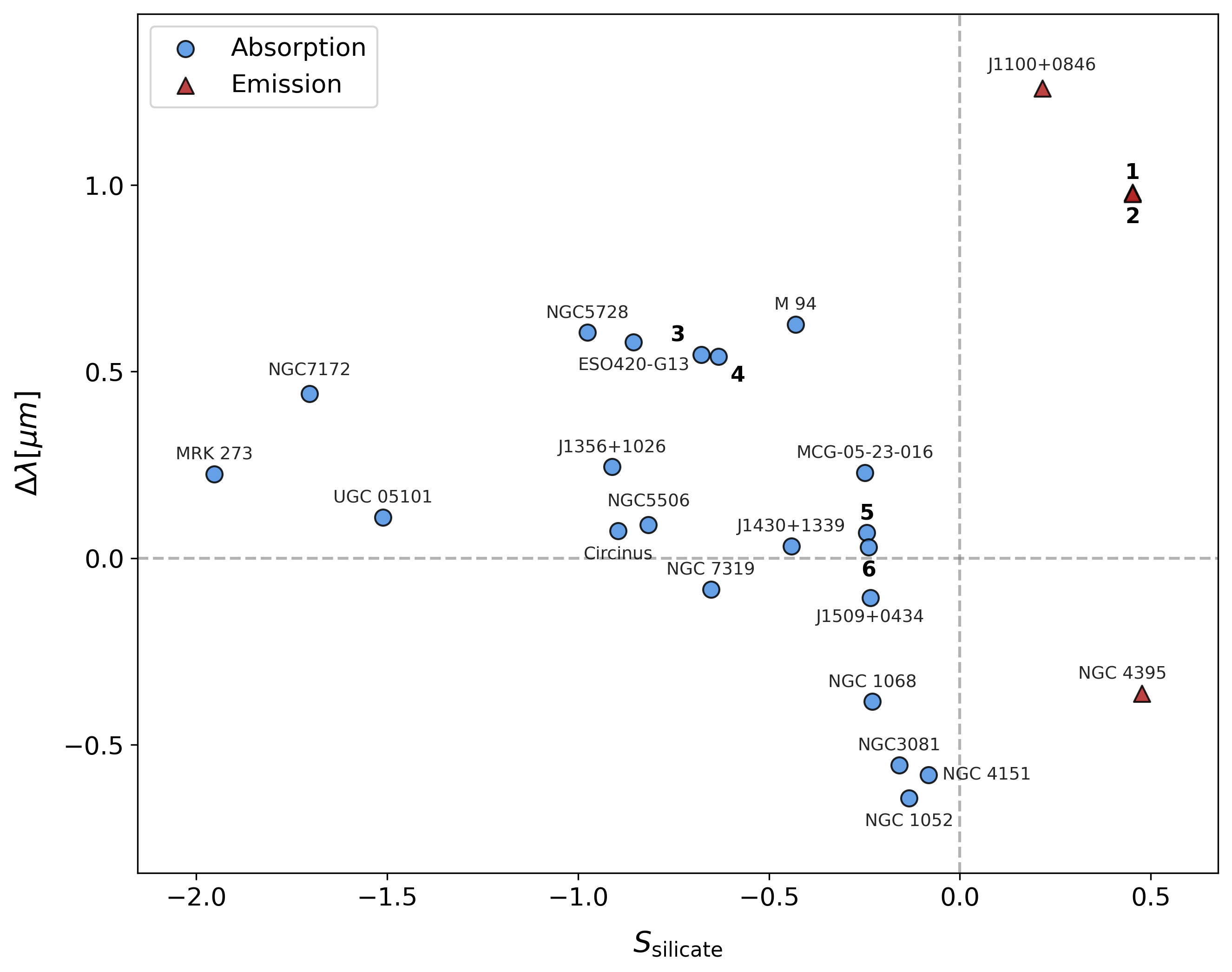}
    \caption{Silicate-feature wavelength displacement, $\Delta\lambda$, versus feature strength, $S_{\rm silicate}$, for the 25 AGN. Blue circles and red triangles denote absorption and emission, respectively. Several crowded sources are numbered: (1) M~81, (2) J1010+0612, (3) NGC~3256, (4) MRK~231, (5) IC~5063, and (6) NGC~7469; all other sources are labelled by name.}
    \label{fig:silicate_shift}
\end{figure*}

To compare the silicate profiles, we measured a characteristic wavelength, $\lambda_{\rm peak}$, and strength, $S_{\rm silicate}=\ln(F_{\rm obs}/F_{\rm cont})$, from the rest-frame, extinction-corrected spectra. For each source, we fitted a local power-law continuum using the same two continuum windows, $7.8$--$8.2$ and $13.5$--$14.5\,\mu\mathrm{m}$, after applying the fixed narrow-line mask. The continuum-normalised profile was then searched between $9.0$ and $11.0\,\mu\mathrm{m}$ to identify the emission maximum or absorption minimum. We define $\Delta\lambda=\lambda_{\rm peak}-9.7\,\mu\mathrm{m}$ and exclude boundary cases from the comparison of peak displacements.

Figure~\ref{fig:silicate_shift} shows that the apparent emission maxima and absorption minima occupy different wavelength ranges. Measurements of weak features near $S_{\rm silicate}=0$ are sensitive to continuum placement and residual PAH emission \citep{2007ApJ...656..770S}. The absorption minima lie from approximately $0.65\,\mu\mathrm{m}$ blueward to $0.45\,\mu\mathrm{m}$ redward of $9.7\,\mu\mathrm{m}$. Four of the five emission-classified sources have maxima displaced redward by $0.8$--$1.3\,\mu\mathrm{m}$; the remaining source, NGC~4395, reaches the blue boundary of the search interval and is therefore not used to infer an emission-peak shift.

Several effects may contribute to this difference. Temperature gradients, self-absorption, and the relative weighting of directly illuminated and obscured regions can shift or broaden the apparent silicate emission profile even when the underlying opacity curve is unchanged \citep{2009ApJ...707.1550N}. Variations in grain size, porosity, and composition may introduce additional changes to the feature profile \citep{2010ApJ...716..490S}. In particular, the preferential survival of larger grains close to the sublimation region is physically plausible because smaller grains are more susceptible to radiative heating and destruction \citep{1993ApJ...402..441L,2020ApJ...892...84T}.

In obscured systems, the observed absorption profile samples a longer and cooler line of sight through the dusty structure. Material at larger radii can therefore contribute strongly near the $9.7\,\mu\mathrm{m}$ silicate resonance, while obscuring the emission from the directly illuminated inner region. This line-of-sight mixing provides a natural qualitative explanation for why the absorption minima remain closer to $9.7\,\mu\mathrm{m}$ than the emission maxima.

Our current model adopts a spatially invariant Astrodust prescription. Extending the calculations to allow spatial variations in grain size or composition will provide a direct way to distinguish grain processing from geometrical and optical-depth effects.

\subsection{Physical implications of the radial density distribution}
\label{subsec:disc_density_physics}

The fiducial grid-score comparison indicates a sample-level preference for shallow radial density distributions: 20 sources select $p=0.5$, one selects $p=1.0$, three select $p=1.5$, and one selects $p=2.0$. The combined weight at $p\leq1$ exceeds 0.5 for 21 sources.

Twelve sources retain the same preferred law across all tested $\beta$, and the number with a majority weight at $p\leq1$ remains between 17 and 21. Varying $\beta$ therefore preserves the sample-level preference for shallow profiles while allowing individual source rankings to change.

For a density law $\rho(r)\propto r^{-p}$, the prevalent shallow profiles imply that the dust density decreases relatively slowly with radius, allowing material beyond the immediate sublimation region to contribute appreciably to the MIR emission. Such an extended distribution is consistent with dusty material being supplied or launched over a broad range of disk radii. Magnetocentrifugal disk winds provide one possible realisation: a coherent poloidal magnetic-field component can extract angular momentum from the disk and lift material along inclined field lines \citep{1982MNRAS.199..883B}. When the dusty wind is fed over an extended radial range, it can form a geometrically and radially distributed obscuring medium \citep{2006ApJ...648L.101E}. Radiative-transfer calculations of disk--wind geometries similarly show that a compact equatorial disk combined with an extended wind can reproduce important properties of AGN infrared emission \citep{2017ApJ...838L..20H}.

A shallow density law is not unique to magnetic launching. Radiative feedback can also lift dusty gas and establish a circulation involving equatorial inflow, high-latitude outflow, and fallback material, producing a geometrically thick and dynamical obscurer on parsec to tens-of-parsec scales \citep{2012ApJ...758...66W}. Radiation-MHD calculations likewise generate extended disk--outflow structures \citep{2017ApJ...843...58C,2019ApJ...876..137W,2020ApJ...897...26W,2020ApJ...900..174V}. The inferred radial density slope can depend on the radial distribution of mass loading, acceleration, and outflow geometry, in addition to the mechanism that launches the material. The result therefore supports an extended, luminosity-weighted MIR-emitting dust distribution, while the launching mechanism remains to be established.

High-angular-resolution observations provide independent motivation for considering radially and vertically extended dust. Parsec-scale interferometry of NGC~424 reveals MIR emission elongated approximately along the polar direction \citep{2012ApJ...755..149H}, while subarcsecond imaging indicates that prominent polar MIR emission is common among nearby Seyfert galaxies \citep{2016ApJ...822..109A,2019MNRAS.489.2177A}. These observations indicate that a substantial fraction of AGN-heated dust can reside outside a purely equatorial torus, consistent with disk--wind or fountain geometries. The tentative increase of the resolved dusty-wind fraction with Eddington ratio also suggests that radiation pressure may contribute to the extended component \citep{2019ApJ...886...55L}.

Magnetic-field measurements offer a complementary test of these scenarios. A magnetocentrifugal wind requires a coherent poloidal field component near its base, although disk rotation can generate a stronger toroidal or helical component farther along the flow \citep{1982MNRAS.199..883B}. Thermal dust polarimetry can constrain the projected magnetic-field morphology, provided that optical-depth effects, grain-alignment physics, and beam averaging are taken into account. ALMA polarimetry of NGC~1068 is consistent with magnetically aligned grains and a dynamically relevant, predominantly toroidal field in the torus \citep{2020ApJ...893...33L}. Joint measurements of MIR source sizes, polar emission fractions, gas kinematics, Eddington ratios, and polarisation will be required to determine whether the inferred density laws reflect different mass-loading regimes or different combinations of magnetic and radiative driving.

\subsection{Constraints on the obscuring geometry and other structural parameters}
\label{sec:discussion_geometry}

The non-radial parameters are less uniquely determined than the density-law ranking. Among the preferred-$p$ MAP solutions, 12 sources lie at the minimum opening angle, 16 at the maximum $Y$, and 11 at the edge-on viewing angle. These concentrations should be interpreted as model-grid preferences rather than direct measurements of a common opening angle, radial extent, or inclination. The effective obscured fraction also depends on the angular density profile, optical depth, and clumpiness \citep{2008ApJ...685..160N,2016MNRAS.458.2288S}.

The component fractions covary with $p$: steeper torus templates require progressively larger SB contributions. This coupling is expected because the spatially extended archive spectra contain host emission and because the two components compete to reproduce both PAH structure and continuum curvature.

The continuum beyond approximately $14\,\mu\mathrm{m}$ requires a larger cool-dust contribution than is available in the present grid. The total dust mass follows from the accretion-linked gas density, $\mathcal{D}$, and the imposed geometry, while raising $\mathcal{D}$ also increases the silicate optical depth. Future grids that vary the cool-dust reservoir more independently will extend the comparison to longer wavelengths.

\subsection{Summary and conclusions}
\label{sec:conclusions}

We constructed a three-dimensional, axisymmetric radiative-transfer library for AGN-heated circumnuclear dust and combined it with empirical starburst templates to model rest-frame $5$--$14\,\mu\mathrm{m}$ MIRI/MRS spectra of 25 AGN. Our goal was to test whether the MIR spectra favour compact or radially extended dust distributions, while also examining the behaviour of the $9.7\,\mu\mathrm{m}$ silicate feature. Our principal results are as follows.

\begin{enumerate}
\item The model reproduces the broad $5$--$14\,\mu\mathrm{m}$ continua and the main PAH and silicate structures for most of the sample. The silicate absorption minima remain near $9.7\,\mu\mathrm{m}$, while four emission-classified sources have maxima shifted redward by approximately $0.8$--$1.3\,\mu\mathrm{m}$. This behaviour is consistent with a combination of radiative-transfer effects and variations in dust properties \citep{2009ApJ...707.1550N,2010ApJ...716..490S}.

\item The normalised grid weights favour shallow radial density laws. Twenty sources select $p=0.5$, one selects $p=1.0$, three select $p=1.5$, and one selects $p=2.0$; 21 have $W(p\leq1\mid D)>0.5$. The result therefore favours a radially extended MIR-emitting dust distribution rather than a strongly centrally concentrated one.

\item The preference for shallow profiles is stable at the sample level. Across the tests with different $\beta$ values, 12 sources retain the same preferred law and the number with a majority weight at $p\leq1$ ranges from 17 to 21. Changing the $\dot m$ prior changes the preferred density law for only one source, indicating that the main density-law result is not driven by the accretion-rate prior.

\item The other structural parameters are less uniquely determined. Within the sampled grid, the preferred solutions tend towards compact openings, large radial extents, and inclined viewing angles, but these values should be interpreted as model-preferred effective parameters rather than direct measurements of a common obscuring geometry.
\end{enumerate}

Taken together, the $5$--$14\,\mu\mathrm{m}$ spectra favour relatively shallow radial dust distributions and show that the observed silicate profiles cannot be interpreted from viewing angle alone. The results support a radially extended MIR-emitting component, broadly consistent with disk-wind or fountain-like obscuring structures. Spatially resolved MIR, molecular-gas, and polarimetric observations will be needed to determine how geometry, dynamics, and dust composition combine to produce the observed spectra.

\section*{Data availability}
The simulated AGN spectral libraries are available on Zenodo at \url{https://doi.org/10.5281/zenodo.18861234}. The processed spectra, fitting scripts and configuration, masks, and final grid summaries used for this revision are provided in the accompanying reproducibility archive.

\begin{acknowledgements}
The authors are grateful to the anonymous referee for their constructive review and insightful suggestions, which significantly improved the clarity and depth of this paper.

This work is based on observations made with the NASA/ESA/CSA James Webb Space Telescope. The data were obtained from the MAST at the Space Telescope Science Institute, which is operated by the Association of Universities for Research in Astronomy, Inc., under NASA contract NAS 5-03127.
\end{acknowledgements}
\bibliographystyle{aa}
\bibliography{reference}

@article{SandersMirabel1996ARAandA34,
 author  = {Sanders, D. B. and Mirabel, I. F.},
 year    = {1996},
 title   = {Luminous Infrared Galaxies},
 journal = {\araa},
 volume  = {34},
 pages   = {749--792}
}

@ARTICLE{2023A&A...676A..73G,
       author = {{Gonz{\'a}lez-Mart{\'\i}n}, Omaira and {Ramos Almeida}, Cristina and {Fritz}, Jacopo and {Alonso-Herrero}, Almudena and {H{\"o}nig}, Sebastian F. and {Roche}, Patrick F. and {Esparza-Arredondo}, Donaji and {Garc{\'\i}a-Bernete}, Ismael and {Garc{\'\i}a-Burillo}, Santiago and {Osorio-Clavijo}, Natalia and {Reyes-Amador}, Ulises and {Stalevski}, Marko and {Victoria-Ceballos}, C{\'e}sar},
        title = "{The role of grain size in active galactic nuclei torus dust models}",
      journal = {\aap},
         year = 2023,
        month = aug,
       volume = {676},
          eid = {A73},
        pages = {A73},
          doi = {10.1051/0004-6361/202345858},
archivePrefix = {arXiv},
       eprint = {2305.11331},
 primaryClass = {astro-ph.GA},
       adsurl = {https://ui.adsabs.harvard.edu/abs/2023A&A...676A..73G}
}

@ARTICLE{2009ApJ...707.1550N,
       author = {{Nikutta}, Robert and {Elitzur}, Moshe and {Lacy}, Mark},
        title = "{On the 10 {\ensuremath{\mu}}m Silicate Feature in Active Galactic Nuclei}",
      journal = {\apj},
         year = 2009,
        month = dec,
       volume = {707},
       number = {2},
        pages = {1550-1559},
          doi = {10.1088/0004-637X/707/2/1550},
archivePrefix = {arXiv},
       eprint = {0910.5521},
 primaryClass = {astro-ph.CO},
       adsurl = {https://ui.adsabs.harvard.edu/abs/2009ApJ...707.1550N}
}

@ARTICLE{2010A&A...523A..27H,
       author = {{H{\"o}nig}, S.~F. and {Kishimoto}, M.},
        title = "{The dusty heart of nearby active galaxies. II. From clumpy torus models to physical properties of dust around AGN}",
      journal = {\aap},
         year = 2010,
        month = nov,
       volume = {523},
          eid = {A27},
        pages = {A27},
          doi = {10.1051/0004-6361/200912676},
archivePrefix = {arXiv},
       eprint = {0909.4539},
 primaryClass = {astro-ph.CO},
       adsurl = {https://ui.adsabs.harvard.edu/abs/2010A&A...523A..27H}
}

@ARTICLE{2008ApJ...685..160N,
       author = {{Nenkova}, Maia and {Sirocky}, Matthew M. and {Nikutta}, Robert and {Ivezi{\'c}}, {\v{Z}}eljko and {Elitzur}, Moshe},
        title = "{AGN Dusty Tori. II. Observational Implications of Clumpiness}",
      journal = {\apj},
         year = 2008,
        month = sep,
       volume = {685},
       number = {1},
        pages = {160-180},
          doi = {10.1086/590483},
archivePrefix = {arXiv},
       eprint = {0806.0512},
 primaryClass = {astro-ph},
       adsurl = {https://ui.adsabs.harvard.edu/abs/2008ApJ...685..160N}
}

@article{DownesSolomon1998ApJ507615,
 author  = {Downes, D. and Solomon, P. M.},
 year    = {1998},
 title   = {Rotating Nuclear Rings and Extreme Starbursts in Ultraluminous Galaxies},
 journal = {\apj},
 volume  = {507},
 pages   = {615--654}
}

@ARTICLE{2007ApJ...663...81P,
       author = {{Polletta}, M. and {Tajer}, M. and {Maraschi}, L. and {Trinchieri}, G. and {Lonsdale}, C.~J. and {Chiappetti}, L. and {Andreon}, S. and {Pierre}, M. and {Le F{\`e}vre}, O. and {Zamorani}, G. and {Maccagni}, D. and {Garcet}, O. and {Surdej}, J. and {Franceschini}, A. and {Alloin}, D. and {Shupe}, D.~L. and {Surace}, J.~A. and {Fang}, F. and {Rowan-Robinson}, M. and {Smith}, H.~E. and {Tresse}, L.},
        title = "{Spectral Energy Distributions of Hard X-Ray Selected Active Galactic Nuclei in the XMM-Newton Medium Deep Survey}",
      journal = {\apj},
         year = 2007,
        month = jul,
       volume = {663},
       number = {1},
        pages = {81-102},
          doi = {10.1086/518113},
archivePrefix = {arXiv},
       eprint = {astro-ph/0703255},
 primaryClass = {astro-ph},
       adsurl = {https://ui.adsabs.harvard.edu/abs/2007ApJ...663...81P}
}

@ARTICLE{1998ApJ...509..103S,
       author = {{Silva}, Laura and {Granato}, Gian Luigi and {Bressan}, Alessandro and {Danese}, Luigi},
        title = "{Modeling the Effects of Dust on Galactic Spectral Energy Distributions from the Ultraviolet to the Millimeter Band}",
      journal = {\apj},
         year = 1998,
        month = dec,
       volume = {509},
       number = {1},
        pages = {103-117},
          doi = {10.1086/306476},
       adsurl = {https://ui.adsabs.harvard.edu/abs/1998ApJ...509..103S}
}

@article{Genzel1998,
 author  = {Genzel, Reinhard and Lutz, Dieter and Sturm, Eckhard and Egami, Eiichi and Kunze, Daniela and Moorwood, Alan F. M. and Rigopoulou, Dimitra and Spoon, Henrik W. W. and Sternberg, Amiel and Tacconi-Garman, Linda E. and Thatte, Niranjan},
 title   = {What Powers Ultraluminous Infrared Galaxies?},
 journal = {\apj},
 year    = {1998},
 volume  = {498},
 pages   = {579--605},
 doi     = {10.1086/305535}
}

@article{Spoon2007,
 author  = {Spoon, H. W. W. and Marshall, J. A. and Houck, J. R. and Elitzur, M. and Hao, L. and Armus, L. and Brandl, B. R. and Charmandaris, V.},
 title   = {Mid-Infrared Galaxy Classification Based on Silicate Obscuration and PAH Equivalent Width},
 journal = {\apjl},
 year    = {2007},
 volume  = {654},
 pages   = {L49--L52},
 doi     = {10.1086/511268}
}

@article{Laurent2000,
 author  = {Laurent, O. and Mirabel, I. F. and Charmandaris, V. and Gallais, P. and Madden, S. C. and Sauvage, M. and Vigroux, L. and Cesarsky, C.},
 title   = {A New Mid-Infrared Diagnostic Diagram for Active Galaxies and Starbursts},
 journal = {\aap},
 year    = {2000},
 volume  = {359},
 pages   = {887--899}
}

@article{Nardini2008,
 author  = {Nardini, E. and Risaliti, G. and Salvati, M. and Sani, E. and Imanishi, M. and Marconi, A.},
 title   = {Disentangling AGN and starburst contributions in ultraluminous infrared galaxies},
 journal = {\mnras},
 year    = {2008},
 volume  = {385},
 pages   = {L130--L134},
 doi     = {10.1111/j.1745-3933.2008.00439.x}
}

@article{ForsterSchreiber2003AA399833,
 author  = {F{\"o}rster Schreiber, N. M. and Genzel, R. and Lutz, D. and Sternberg, A. and others},
 year    = {2003},
 title   = {The Starburst in {M\,82}: Mid-infrared Spectroscopy and Diagnostics},
 journal = {\aap},
 volume  = {399},
 pages   = {833--852}
}

@ARTICLE{2001A&A...377...73R,
       author = {{Radovich}, M. and {Kahanp{\"a}{\"a}}, J. and {Lemke}, D.},
        title = "{Far-infrared mapping of the starburst galaxy <ASTROBJ>NGC 253</ASTROBJ> with ISOPHOT}",
      journal = {\aap},
         year = 2001,
        month = oct,
       volume = {377},
        pages = {73-83},
          doi = {10.1051/0004-6361:20011102},
archivePrefix = {arXiv},
       eprint = {astro-ph/0108075},
 primaryClass = {astro-ph},
       adsurl = {https://ui.adsabs.harvard.edu/abs/2001A&A...377...73R}
}

@article{Dale2012ApJ74595,
 author  = {Dale, D. A. and Aniano, G. and Engelbracht, C. W. and others},
 year    = {2012},
 title   = {The Spectral Energy Distributions of Local Galaxies from 3 to 500~$\mu$m: Analysis of the {KINGFISH} Sample},
 journal = {\apj},
 volume  = {745},
 pages   = {95}
}

@ARTICLE{2006ApJ...653.1129B,
       author = {{Brandl}, B.~R. and {Bernard-Salas}, J. and {Spoon}, H.~W.~W. and {Devost}, D. and {Sloan}, G.~C. and {Guilles}, S. and {Wu}, Y. and {Houck}, J.~R. and {Weedman}, D.~W. and {Armus}, L. and {Appleton}, P.~N. and {Soifer}, B.~T. and {Charmandaris}, V. and {Hao}, L. and {Higdon}, J.~A. and {Marshall}, S.~J. and {Herter}, T.~L.},
        title = "{The Mid-Infrared Properties of Starburst Galaxies from Spitzer-IRS Spectroscopy}",
      journal = {\apj},
         year = 2006,
        month = dec,
       volume = {653},
       number = {2},
        pages = {1129-1144},
          doi = {10.1086/508849},
archivePrefix = {arXiv},
       eprint = {astro-ph/0609024},
 primaryClass = {astro-ph},
       adsurl = {https://ui.adsabs.harvard.edu/abs/2006ApJ...653.1129B}
}

@article{Antonucci1993,
 title={Unified models for active galactic nuclei and quasars},
 author={Antonucci, Robert},
 journal={\araa},
 volume={31},
 number={1},
 pages={473--521},
 year={1993},
 publisher={Annual Reviews}
}

@ARTICLE{2005ApJ...631..689E,
       author = {{Everett}, John E.},
        title = "{Radiative Transfer and Acceleration in Magnetocentrifugal Winds}",
      journal = {\apj},
         year = 2005,
        month = oct,
       volume = {631},
       number = {2},
        pages = {689-706},
          doi = {10.1086/432678},
archivePrefix = {arXiv},
       eprint = {astro-ph/0506321},
 primaryClass = {astro-ph},
       adsurl = {https://ui.adsabs.harvard.edu/abs/2005ApJ...631..689E}
}

@article{Contopoulos1994,
 author  = {Contopoulos, J. and Lovelace, R. V. E.},
 title   = {Magnetized Accretion-Disk Winds},
 journal = {\apj},
 year    = {1994},
 volume  = {429},
 pages   = {139}
}

@article{Behar2009,
 author  = {Behar, E. and others},
 title   = {X-ray Absorption in MCG-6-30-15},
 journal = {\apj},
 year    = {2009},
 volume  = {703},
 pages   = {1347}
}

@ARTICLE{2025MNRAS.539.2158G,
       author = {{Gonz{\'a}lez-Mart{\'\i}n}, Omaira and {D{\'\i}az-Gonz{\'a}lez}, Daniel J. and {Mart{\'\i}nez-Paredes}, Mariela and {Alonso-Herrero}, Almudena and {L{\'o}pez-Rodr{\'\i}guez}, Enrique and {Garc{\'\i}a-Lorenzo}, Bego{\~n}a and {Ramos Almeida}, Cristina and {Garc{\'\i}a-Bernete}, Ismael and {Esparza-Arredondo}, Donaji and {Hoenig}, Sebastian F. and {Garc{\'\i}a-Burillo}, Santiago and {Packham}, Chris and {Levenson}, Nancy A. and {Labiano}, Alvaro and {Pereira-Santaella}, Miguel and {Combes}, Francoise and {Audibert}, Anelise and {Hicks}, Erin K.~S. and {Zhang}, Lulu and {Bellocchi}, Enrica and {Davies}, Richard I. and {Mu{\~n}oz}, Laura Hermosa and {Imanishi}, Masatoshi and {Ricci}, Claudio and {Stalevski}, Marko},
        title = "{JWST reveals the diversity of nuclear obscuring dust in nearby AGN: nuclear isolation of MIRI/MRS data cubes and continuum spectral fitting}",
      journal = {\mnras},
         year = 2025,
        month = may,
       volume = {539},
       number = {3},
        pages = {2158-2184},
          doi = {10.1093/mnras/staf573},
archivePrefix = {arXiv},
       eprint = {2504.01103},
 primaryClass = {astro-ph.GA},
       adsurl = {https://ui.adsabs.harvard.edu/abs/2025MNRAS.539.2158G}
}

@article{Valiante2011,
 author  = {Valiante, R. and others},
 title   = {Dust Production in the Early Universe},
 journal = {\mnras},
 year    = {2011},
 volume  = {416},
 pages   = {1916--1932}
}

@ARTICLE{2012ApJ...755..149H,
       author = {{H{\"o}nig}, S.~F. and {Kishimoto}, M. and {Antonucci}, R. and {Marconi}, A. and {Prieto}, M.~A. and {Tristram}, K. and {Weigelt}, G.},
        title = "{Parsec-scale Dust Emission from the Polar Region in the Type 2 Nucleus of NGC 424}",
      journal = {\apj},
         year = 2012,
        month = aug,
       volume = {755},
       number = {2},
          eid = {149},
        pages = {149},
          doi = {10.1088/0004-637X/755/2/149},
archivePrefix = {arXiv},
       eprint = {1206.4307},
 primaryClass = {astro-ph.CO},
       adsurl = {https://ui.adsabs.harvard.edu/abs/2012ApJ...755..149H}
}

@ARTICLE{2016ApJ...822..109A,
       author = {{Asmus}, D. and {H{\"o}nig}, S.~F. and {Gandhi}, P.},
        title = "{The Subarcsecond Mid-infrared View of Local Active Galactic Nuclei. III. Polar Dust Emission}",
      journal = {\apj},
         year = 2016,
        month = may,
       volume = {822},
       number = {2},
          eid = {109},
        pages = {109},
          doi = {10.3847/0004-637X/822/2/109},
archivePrefix = {arXiv},
       eprint = {1603.02710},
 primaryClass = {astro-ph.GA},
       adsurl = {https://ui.adsabs.harvard.edu/abs/2016ApJ...822..109A}
}

@BOOK{1979rpa..book.....R,
  author    = {{Rybicki}, George B. and {Lightman}, Alan P.},
  title     = {Radiative Processes in Astrophysics},
  year      = {1979},
  publisher = {Wiley},
  address   = {New York}
}

@article{Gamez-Rosas2023,
 author  = {Gamez-Rosas, V. and others},
 title   = {Prospects for TMT/MICHI Observations},
 journal = {\pasp},
 year    = {2023},
 volume  = {135},
 pages   = {068001}
}

@article{Pier1992,
 title={The dusty torus in active galactic nuclei},
 author={Pier, Edward A and Krolik, Julian H},
 journal={\apj},
 volume={401},
 pages={99},
 year={1992},
 publisher={IOP Publishing}
}

@article{Fritz2006,
 title={The AGN dusty torus: A unified view},
 author={Fritz, Jacopo and Franceschini, Alberto and Hatziminaoglou, Evanthia},
 journal={\mnras},
 volume={366},
 number={2},
 pages={767--786},
 year={2006},
 publisher={Oxford University Press}
}

@article{Nenkova2008,
 title={Clumpy tori in AGN},
 author={Nenkova, Maia and Sirocky, Matthew M and Ivezić, {\v Z}eljko and Elitzur, Moshe},
 journal={\apj},
 volume={685},
 number={1},
 pages={147},
 year={2008},
 publisher={IOP Publishing}
}

@article{Hensley_2023,
    title={The Astrodust+PAH Model: A Unified Description of the Extinction, Emission, and Polarization from Dust in the Diffuse Interstellar Medium},
    volume={948},
    ISSN={1538-4357},
    url={http://dx.doi.org/10.3847/1538-4357/acc4c2},
    DOI={10.3847/1538-4357/acc4c2},
    number={1},
    journal={\apj},
    publisher={American Astronomical Society},
    author={Hensley, Brandon S. and Draine, B. T.},
    year={2023},
    month=may, pages={55} }

@article{Draine2007,
 author  = {Draine, B. T. and Li, A.},
 title   = {Radiation-Driven Dust Models: {PAHs}, Silicates, and the {ISM}},
 journal = {\apj},
 year    = {2007},
 volume  = {657},
 pages   = {810-837},
 doi     = {10.1086/521824}
}

@article{Sarangi_2019,
    title={Dust Formation in AGN Winds},
    volume={885},
    ISSN={1538-4357},
    url={http://dx.doi.org/10.3847/1538-4357/ab46a9},
    DOI={10.3847/1538-4357/ab46a9},
    number={2},
    journal={\apj},
    publisher={American Astronomical Society},
    author={Sarangi, Arkaprabha and Dwek, Eli and Kazanas, Demos},
    year={2019},
    month=nov, pages={126} }

@MISC{AGN,
  author = {{Ricci}, C.},
  title  = {The Typical AGN Is Made of Several Components},
  year   = {n.d.},
  note   = {\url{https://www.isdc.unige.ch/~ricci/Website/Active_Galactic_Nuclei.html}}
}

@article{Stalevski_2012,
    title={3D radiative transfer modelling of the dusty tori around active galactic nuclei as a clumpy two-phase medium: AGN dusty tori as clumpy two-phase medium},
    volume={420},
    ISSN={0035-8711},
    url={http://dx.doi.org/10.1111/j.1365-2966.2011.19775.x},
    DOI={10.1111/j.1365-2966.2011.19775.x},
    number={4},
    journal={\mnras},
    publisher={Oxford University Press (OUP)},
    author={Stalevski, Marko and Fritz, Jacopo and Baes, Maarten and Nakos, Theodoros and Popović, Luka {\v C}.},
    year={2012},
    month=feb, pages={2756–2772} }

@ARTICLE{1977ApJ...217..425M,
       author = {{Mathis}, J.~S. and {Rumpl}, W. and {Nordsieck}, K.~H.},
        title = "{The size distribution of interstellar grains.}",
      journal = {\apj},
         year = 1977,
        month = oct,
       volume = {217},
        pages = {425-433},
          doi = {10.1086/155591},
       adsurl = {https://ui.adsabs.harvard.edu/abs/1977ApJ...217..425M}
}

@article{Antonucci1985,
 author = {Antonucci, R. R. J. and Miller, J. S.},
 title = {Spectropolarimetry and the Nature of NGC 1068},
 journal = {\apj},
 year = {1985},
 volume = {297},
 pages = {621--632},
 doi = {10.1086/163561}
}

@article{Dwek1987,
 author = {Dwek, E.},
 title = {The evolution of refractory interstellar grains},
 journal = {\apj},
 volume = {322},
 pages = {812--821},
 year = {1987}
}

@article{Fukumura2010,
 author = {Fukumura, K. and Kazanas, D. and Contopoulos, I. and Behar, E.},
 title = {Magnetohydrodynamic Accretion-Disk Winds as X-ray Absorbers in Active Galactic Nuclei},
 journal = {\apj},
 volume = {715},
 pages = {636--650},
 year = {2010}
}

@article{Lopez-Gonzaga2016,
 author = {López-Gonzaga, N. and Jaffe, W. J.},
 title = {VLTI/MIDI Atlas of AGN Dust Tori: Polar-elongated Mid-infrared Emission},
 journal = {\mnras},
 year = {2016},
 volume = {462},
 pages = {4183--4195},
 doi = {10.1093/mnras/stw1911}
}

@article{Draine_2021,
    title={The Dielectric Function of ''Astrodus'' and Predictions for Polarization in the 3.4 and 10 μm Features},
    volume={909},
    ISSN={1538-4357},
    url={http://dx.doi.org/10.3847/1538-4357/abd6c6},
    DOI={10.3847/1538-4357/abd6c6},
    number={1},
    journal={\apj},
    publisher={American Astronomical Society},
    author={Draine, B. T. and Hensley, Brandon S.},
    year={2021},
    month=mar, pages={94} }

@ARTICLE{2000ApJ...545...63E,
       author = {{Elvis}, Martin},
        title = "{A Structure for Quasars}",
      journal = {\apj},
         year = 2000,
        month = dec,
       volume = {545},
       number = {1},
        pages = {63-76},
          doi = {10.1086/317778},
archivePrefix = {arXiv},
       eprint = {astro-ph/0008064},
 primaryClass = {astro-ph},
       adsurl = {https://ui.adsabs.harvard.edu/abs/2000ApJ...545...63E}
}

@ARTICLE{Remy-Ruyer2014,
  author  = {{R{\'e}my-Ruyer}, A. and {Madden}, S. C. and {Galliano}, F. and others},
  title   = {Gas-to-dust mass ratios in local galaxies over a 2 dex metallicity range},
  journal = {\aap},
  year    = {2014},
  volume  = {563},
  eid     = {A13},
  pages   = {A13},
  doi     = {10.1051/0004-6361/201322803}
}

@ARTICLE{1987MNRAS.225...55N,
       author = {{Netzer}, Hagai},
        title = "{Quasar discs. II - A composite model for the broad-line region}",
      journal = {\mnras},
         year = 1987,
        month = mar,
       volume = {225},
        pages = {55-72},
          doi = {10.1093/mnras/225.1.55},
       adsurl = {https://ui.adsabs.harvard.edu/abs/1987MNRAS.225...55N}
}

@article{Shakura1973,
    author = {Shakura, N. I. and Sunyaev, R. A.},
    title = "{Black holes in binary systems. Observational appearance.}",
    journal = {\aap},
    volume = {24},
    pages = {337-355},
    year = {1973},
}

@article{Haardt1993,
    author = {Haardt, Francesco and Maraschi, Laura},
    title = "{A two-phase model for the X-ray emission from Seyfert galaxies}",
    journal = {\apj},
    volume = {413},
    pages = {507-517},
    year = {1993},
    doi = {10.1086/173020},
}

@article{Lusso2016,
    author = {Lusso, E. and Worseck, G. and Hennawi, J. F. and Prochaska, J. X. and Vignali, C. and Stern, J. and O'Meara, J. M.},
    title = "{The first X-ray survey of optically selected quasars with the SRG/eROSITA}",
    journal = {\mnras},
    volume = {456},
    pages = {1253-1268},
    year = {2016},
    doi = {10.1093/mnras/stv2737},
}

@ARTICLE{2021ApJ...916...90B,
       author = {{Balokovi{\'c}}, M. and {Cabral}, S.~E. and {Brenneman}, L. and {Urry}, C.~M.},
        title = "{Properties of the Obscuring Torus in NGC 1052 from Multiepoch Broadband X-Ray Spectroscopy}",
      journal = {\apj},
         year = 2021,
        month = aug,
       volume = {916},
       number = {2},
          eid = {90},
        pages = {90},
          doi = {10.3847/1538-4357/abff4d},
archivePrefix = {arXiv},
       eprint = {2105.01682},
 primaryClass = {astro-ph.HE},
       adsurl = {https://ui.adsabs.harvard.edu/abs/2021ApJ...916...90B}
}

@ARTICLE{2020MNRAS.497.1020W,
       author = {{Wang}, Jian-Min and {Songsheng}, Yu-Yang and {Li}, Yan-Rong and {Du}, Pu and {Yu}, Zhe},
        title = "{Dynamical evidence from the sub-parsec counter-rotating disc for a close binary of supermassive black holes in NGC 1068}",
      journal = {\mnras},
         year = 2020,
        month = sep,
       volume = {497},
       number = {1},
        pages = {1020-1028},
          doi = {10.1093/mnras/staa1985},
archivePrefix = {arXiv},
       eprint = {2005.01220},
 primaryClass = {astro-ph.GA},
       adsurl = {https://ui.adsabs.harvard.edu/abs/2020MNRAS.497.1020W}
}

@ARTICLE{2009ApJ...698..528B,
       author = {{Brenneman}, L.~W. and {Weaver}, K.~A. and {Kadler}, M. and {Tueller}, J. and {Marscher}, A. and {Ros}, E. and {Zensus}, A. and {Kovalev}, Y.~Y. and {Aller}, M. and {Aller}, H. and {Irwin}, J. and {Kerp}, J. and {Kaufmann}, S.},
        title = "{Spectral Analysis of the Accretion Flow in NGC 1052 with Suzaku}",
      journal = {\apj},
         year = 2009,
        month = jun,
       volume = {698},
       number = {1},
        pages = {528-540},
          doi = {10.1088/0004-637X/698/1/528},
archivePrefix = {arXiv},
       eprint = {0903.3583},
 primaryClass = {astro-ph.HE},
       adsurl = {https://ui.adsabs.harvard.edu/abs/2009ApJ...698..528B}
}

@ARTICLE{2024ApJ...974..195Z,
       author = {{Zhang}, Lulu and {Packham}, Chris and {Hicks}, Erin K.~S. and {Davies}, Ric I. and {Shimizu}, Taro T. and {Alonso-Herrero}, Almudena and {Hermosa Mu{\~n}oz}, Laura and {Garc{\'\i}a-Bernete}, Ismael and {Pereira-Santaella}, Miguel and {Audibert}, Anelise and {L{\'o}pez-Rodr{\'\i}guez}, Enrique and {Bellocchi}, Enrica and {Bunker}, Andrew J. and {Combes}, Francoise and {D{\'\i}az-Santos}, Tanio and {Gandhi}, Poshak and {Garc{\'\i}a-Burillo}, Santiago and {Garc{\'\i}a-Lorenzo}, Bego{\~n}a and {Gonz{\'a}lez-Mart{\'\i}n}, Omaira and {Imanishi}, Masatoshi and {Labiano}, Alvaro and {Leist}, Mason T. and {Levenson}, Nancy A. and {Ramos Almeida}, Cristina and {Ricci}, Claudio and {Rigopoulou}, Dimitra and {Rosario}, David J. and {Stalevski}, Marko and {Ward}, Martin J. and {Esparza-Arredondo}, Donaji and {Delaney}, Dan and {Fuller}, Lindsay and {Haidar}, Houda and {H{\"o}nig}, Sebastian and {Izumi}, Takuma and {Rouan}, Daniel},
        title = "{The Galaxy Activity, Torus, and Outflow Survey (GATOS). IV. Exploring Ionized Gas Outflows in Central Kiloparsec Regions of GATOS Seyferts}",
      journal = {\apj},
         year = 2024,
        month = oct,
       volume = {974},
       number = {2},
          eid = {195},
        pages = {195},
          doi = {10.3847/1538-4357/ad6a4b},
archivePrefix = {arXiv},
       eprint = {2409.09771},
 primaryClass = {astro-ph.GA},
       adsurl = {https://ui.adsabs.harvard.edu/abs/2024ApJ...974..195Z}
}

@ARTICLE{2020633A.127F,
       author = {{Fern{\'a}ndez-Ontiveros}, J.~A. and {Dasyra}, K.~M. and {Hatziminaoglou}, E. and {Malkan}, M.~A. and {Pereira-Santaella}, M. and {Papachristou}, M. and {Spinoglio}, L. and {Combes}, F. and {Aalto}, S. and {Nagar}, N. and {Imanishi}, M. and {Andreani}, P. and {Ricci}, C. and {Slater}, R.},
        title = "{A CO molecular gas wind 340 pc away from the Seyfert 2 nucleus in ESO 420-G13 probes an elusive radio jet}",
      journal = {\aap},
         year = 2020,
        month = jan,
       volume = {633},
          eid = {A127},
        pages = {A127},
          doi = {10.1051/0004-6361/201936552},
archivePrefix = {arXiv},
       eprint = {1911.00015},
 primaryClass = {astro-ph.GA},
       adsurl = {https://ui.adsabs.harvard.edu/abs/2020A&A...633A.127F}
}

@ARTICLE{2003AJ....125.1226D,
       author = {{Devereux}, Nick and {Ford}, Holland and {Tsvetanov}, Zlatan and {Jacoby}, George},
        title = "{STIS Spectroscopy of the Central 10 Parsecs of M81: Evidence for a Massive Black Hole}",
      journal = {\aj},
         year = 2003,
        month = mar,
       volume = {125},
       number = {3},
        pages = {1226-1235},
          doi = {10.1086/367595},
       adsurl = {https://ui.adsabs.harvard.edu/abs/2003AJ....125.1226D}
}

@ARTICLE{2012542A..83P,
       author = {{Ponti}, G. and {Papadakis}, I. and {Bianchi}, S. and {Guainazzi}, M. and {Matt}, G. and {Uttley}, P. and {Bonilla}, N.~F.},
        title = "{CAIXA: a catalogue of AGN in the XMM-Newton archive. III. Excess variance analysis}",
      journal = {\aap},
         year = 2012,
        month = jun,
       volume = {542},
          eid = {A83},
        pages = {A83},
          doi = {10.1051/0004-6361/201118326},
archivePrefix = {arXiv},
       eprint = {1112.2744},
 primaryClass = {astro-ph.HE},
       adsurl = {https://ui.adsabs.harvard.edu/abs/2012A&A...542A..83P}
}

@ARTICLE{2009ApJ...692..856B,
       author = {{Beifiori}, A. and {Sarzi}, M. and {Corsini}, E.~M. and {Dalla Bont{\`a}}, E. and {Pizzella}, A. and {Coccato}, L. and {Bertola}, F.},
        title = "{Upper Limits on the Masses of 105 Supermassive Black Holes from Hubble Space Telescope/Space Telescope Imaging Spectrograph Archival Data}",
      journal = {\apj},
         year = 2009,
        month = feb,
       volume = {692},
       number = {1},
        pages = {856-868},
          doi = {10.1088/0004-637X/692/1/856},
archivePrefix = {arXiv},
       eprint = {0809.5103},
 primaryClass = {astro-ph},
       adsurl = {https://ui.adsabs.harvard.edu/abs/2009ApJ...692..856B}
}

@ARTICLE{2013ApJ...765...78A,
       author = {{Alonso-Herrero}, Almudena and {Pereira-Santaella}, Miguel and {Rieke}, George H. and {Diamond-Stanic}, Aleksandar M. and {Wang}, Yiping and {Hern{\'a}n-Caballero}, Antonio and {Rigopoulou}, Dimitra},
        title = "{Local Luminous Infrared Galaxies. III. Co-evolution of Black Hole Growth and Star Formation Activity?}",
      journal = {\apj},
         year = 2013,
        month = mar,
       volume = {765},
       number = {2},
          eid = {78},
        pages = {78},
          doi = {10.1088/0004-637X/765/2/78},
archivePrefix = {arXiv},
       eprint = {1301.4015},
 primaryClass = {astro-ph.CO},
       adsurl = {https://ui.adsabs.harvard.edu/abs/2013ApJ...765...78A}
}

@article{Brok_2015,
   title={MEASURING THE MASS OF THE CENTRAL BLACK HOLE IN THE BULGELESS GALAXY NGC 4395 FROM GAS DYNAMICAL MODELING},
   volume={809},
   ISSN={1538-4357},
   url={http://dx.doi.org/10.1088/0004-637X/809/1/101},
   DOI={10.1088/0004-637x/809/1/101},
   number={1},
   journal={ApJ},
   publisher={American Astronomical Society},
   author={Brok, Mark den and Seth, Anil C. and Barth, Aaron J. and Carson, Daniel J. and Neumayer, Nadine and Cappellari, Michele and Debattista, Victor P. and Ho, Luis C. and Hood, Carol E. and McDermid, Richard M.},
   year={2015},
   month=aug, pages={101} }

@ARTICLE{cruz2023modelingsedagninside,
      title={Modeling the SED of the AGN inside NGC 4395}, 
      author={Hector Afonso G. Cruz and Andy D. Goulding and Jenny E. Greene},
      year={2023},
      eprint={2307.15111},
      archivePrefix={arXiv},
      primaryClass={astro-ph.GA}, journal={arXiv e-prints}, eid={arXiv:2307.15111},
      url={https://arxiv.org/abs/2307.15111}, 
}

@ARTICLE{2025AA...698L...9N,
       author = {{Nguyen}, Dieu D. and {Ngo}, Hai N. and {Le}, Tinh Q.~T. and {Graham}, Alister W. and {Soria}, Roberto and {Chilingarian}, Igor V. and {Thatte}, Niranjan and {Phuong}, N.~T. and {Hoang}, Thiem and {Pereira-Santaella}, Miguel and {Durre}, Mark and {Pham}, Diep N. and {Ngoc Tram}, Le and {Ngoc}, Nguyen B. and {L{\^e}}, Ng{\^a}n},
        title = "{Supermassive black hole mass measurement in the spiral galaxy NGC 4736 using JWST/NIRSpec stellar kinematics}",
      journal = {\aap},
         year = 2025,
        month = jun,
       volume = {698},
          eid = {L9},
        pages = {L9},
          doi = {10.1051/0004-6361/202554672},
archivePrefix = {arXiv},
       eprint = {2505.09941},
 primaryClass = {astro-ph.GA},
       adsurl = {https://ui.adsabs.harvard.edu/abs/2025A&A...698L...9N}
}

@article{Yan_2015,
   title={A PROBABLE MILLI-PARSEC SUPERMASSIVE BINARY BLACK HOLE IN THE NEAREST QUASAR MRK 231},
   volume={809},
   ISSN={1538-4357},
   url={http://dx.doi.org/10.1088/0004-637X/809/2/117},
   DOI={10.1088/0004-637x/809/2/117},
   number={2},
   journal={ApJ},
   publisher={American Astronomical Society},
   author={Yan, Chang-Shuo and Lu, Youjun and Dai, Xinyu and Yu, Qingjuan},
   year={2015},
   month=aug, pages={117} }

@ARTICLE{2013ApJ...775..115U,
       author = {{U}, Vivian and {Medling}, Anne and {Sanders}, David and {Max}, Claire and {Armus}, Lee and {Iwasawa}, Kazushi and {Evans}, Aaron and {Kewley}, Lisa and {Fazio}, Giovanni},
        title = "{The Inner Kiloparsec of Mrk 273 with Keck Adaptive Optics}",
      journal = {\apj},
         year = 2013,
        month = oct,
       volume = {775},
       number = {2},
          eid = {115},
        pages = {115},
          doi = {10.1088/0004-637X/775/2/115},
archivePrefix = {arXiv},
       eprint = {1307.8440},
 primaryClass = {astro-ph.CO},
       adsurl = {https://ui.adsabs.harvard.edu/abs/2013ApJ...775..115U}
}

@article{Niko_ajuk_2009,
   title={NLS1 galaxies and estimation of their central black hole masses from the X-ray excess variance method},
   volume={394},
   ISSN={1365-2966},
   url={http://dx.doi.org/10.1111/j.1365-2966.2009.14478.x},
   DOI={10.1111/j.1365-2966.2009.14478.x},
   number={4},
   journal={MNRAS},
   publisher={Oxford University Press (OUP)},
   author={Nikołajuk, M. and Czerny, B. and Gurynowicz, P.},
   year={2009},
   month=apr, pages={2141–2152} }

@ARTICLE{2022AA...666A.127A,
       author = {{Akylas}, A. and {Papadakis}, I. and {Georgakakis}, A.},
        title = "{Black hole mass estimation using X-ray variability measurements in Seyfert galaxies}",
      journal = {\aap},
         year = 2022,
        month = oct,
       volume = {666},
          eid = {A127},
        pages = {A127},
          doi = {10.1051/0004-6361/202244162},
archivePrefix = {arXiv},
       eprint = {2208.12490},
 primaryClass = {astro-ph.HE},
       adsurl = {https://ui.adsabs.harvard.edu/abs/2022A&A...666A.127A}
}

@ARTICLE{2019ApJ...870...37D,
       author = {{Durr{\'e}}, Mark and {Mould}, Jeremy},
        title = "{The AGN Ionization Cones of NGC 5728. II. Kinematics}",
      journal = {\apj},
         year = 2019,
        month = jan,
       volume = {870},
       number = {1},
          eid = {37},
        pages = {37},
          doi = {10.3847/1538-4357/aaf000},
archivePrefix = {arXiv},
       eprint = {1811.04513},
 primaryClass = {astro-ph.GA},
       adsurl = {https://ui.adsabs.harvard.edu/abs/2019ApJ...870...37D}
}

@ARTICLE{2012ApJ...748..130M,
       author = {{Marinucci}, Andrea and {Bianchi}, Stefano and {Nicastro}, Fabrizio and {Matt}, Giorgio and {Goulding}, Andy D.},
        title = "{The Link between the Hidden Broad Line Region and the Accretion Rate in Seyfert 2 Galaxies}",
      journal = {\apj},
         year = 2012,
        month = apr,
       volume = {748},
       number = {2},
          eid = {130},
        pages = {130},
          doi = {10.1088/0004-637X/748/2/130},
archivePrefix = {arXiv},
       eprint = {1201.5397},
 primaryClass = {astro-ph.CO},
       adsurl = {https://ui.adsabs.harvard.edu/abs/2012ApJ...748..130M}
}

@ARTICLE{2022ApJ...936L..14P,
       author = {{Pontoppidan}, Klaus M. and {Barrientes}, Jaclyn and {Blome}, Claire and {Braun}, Hannah and {Brown}, Matthew and {Carruthers}, Margaret and {Coe}, Dan and {DePasquale}, Joseph and {Espinoza}, N{\'e}stor and {Marin}, Macarena Garcia and {Gordon}, Karl D. and {Henry}, Alaina and {Hustak}, Leah and {James}, Andi and {Jenkins}, Ann and {Koekemoer}, Anton M. and {LaMassa}, Stephanie and {Law}, David and {Lockwood}, Alexandra and {Moro-Martin}, Amaya and {Mullally}, Susan E. and {Pagan}, Alyssa and {Player}, Dani and {Proffitt}, Charles and {Pulliam}, Christine and {Ramsay}, Leah and {Ravindranath}, Swara and {Reid}, Neill and {Robberto}, Massimo and {Sabbi}, Elena and {Ubeda}, Leonardo and {Balogh}, Michael and {Flanagan}, Kathryn and {Gardner}, Jonathan and {Hasan}, Hashima and {Meinke}, Bonnie and {Nota}, Antonella},
        title = "{The JWST Early Release Observations}",
      journal = {\apjl},
         year = 2022,
        month = sep,
       volume = {936},
       number = {1},
          eid = {L14},
        pages = {L14},
          doi = {10.3847/2041-8213/ac8a4e},
archivePrefix = {arXiv},
       eprint = {2207.13067},
 primaryClass = {astro-ph.IM},
       adsurl = {https://ui.adsabs.harvard.edu/abs/2022ApJ...936L..14P}
}

@ARTICLE{2014SSRv..183..253P,
       author = {{Peterson}, Bradley M.},
        title = "{Measuring the Masses of Supermassive Black Holes}",
      journal = {\ssr},
         year = 2014,
        month = sep,
       volume = {183},
       number = {1-4},
        pages = {253-275},
          doi = {10.1007/s11214-013-9987-4},
       adsurl = {https://ui.adsabs.harvard.edu/abs/2014SSRv..183..253P}
}

@ARTICLE{2021ApJ...916...25R,
       author = {{Roberts}, Caroline A. and {Bentz}, Misty C. and {Vasiliev}, Eugene and {Valluri}, Monica and {Onken}, Christopher A.},
        title = "{The Black Hole Mass of NGC 4151 from Stellar Dynamical Modeling}",
      journal = {\apj},
         year = 2021,
        month = jul,
       volume = {916},
       number = {1},
          eid = {25},
        pages = {25},
          doi = {10.3847/1538-4357/ac05b6},
archivePrefix = {arXiv},
       eprint = {2106.02758},
 primaryClass = {astro-ph.GA},
       adsurl = {https://ui.adsabs.harvard.edu/abs/2021ApJ...916...25R}
}

@ARTICLE{2017ApJ...850...74K,
       author = {{Koss}, Michael and {Trakhtenbrot}, Benny and {Ricci}, Claudio and {Lamperti}, Isabella and {Oh}, Kyuseok and {Berney}, Simon and {Schawinski}, Kevin and {Balokovi{\'c}}, Mislav and {Baronchelli}, Linda and {Crenshaw}, D. Michael and {Fischer}, Travis and {Gehrels}, Neil and {Harrison}, Fiona and {Hashimoto}, Yasuhiro and {Hogg}, Drew and {Ichikawa}, Kohei and {Masetti}, Nicola and {Mushotzky}, Richard and {Sartori}, Lia and {Stern}, Daniel and {Treister}, Ezequiel and {Ueda}, Yoshihiro and {Veilleux}, Sylvain and {Winter}, Lisa},
        title = "{BAT AGN Spectroscopic Survey. I. Spectral Measurements, Derived Quantities, and AGN Demographics}",
      journal = {\apj},
         year = 2017,
        month = nov,
       volume = {850},
       number = {1},
          eid = {74},
        pages = {74},
          doi = {10.3847/1538-4357/aa8ec9},
archivePrefix = {arXiv},
       eprint = {1707.08123},
 primaryClass = {astro-ph.HE},
       adsurl = {https://ui.adsabs.harvard.edu/abs/2017ApJ...850...74K}
}

@ARTICLE{2021ApJS..257...61Y,
       author = {{Yamada}, Satoshi and {Ueda}, Yoshihiro and {Tanimoto}, Atsushi and {Imanishi}, Masatoshi and {Toba}, Yoshiki and {Ricci}, Claudio and {Privon}, George C.},
        title = "{Comprehensive Broadband X-Ray and Multiwavelength Study of Active Galactic Nuclei in 57 Local Luminous and Ultraluminous Infrared Galaxies Observed with NuSTAR and/or Swift/BAT}",
      journal = {\apjs},
         year = 2021,
        month = dec,
       volume = {257},
       number = {2},
          eid = {61},
        pages = {61},
          doi = {10.3847/1538-4365/ac17f5},
archivePrefix = {arXiv},
       eprint = {2107.10855},
 primaryClass = {astro-ph.GA},
       adsurl = {https://ui.adsabs.harvard.edu/abs/2021ApJS..257...61Y}
}

@ARTICLE{2018ApJ...859..116K,
       author = {{Kong}, Minzhi and {Ho}, Luis C.},
        title = "{The Black Hole Masses and Eddington Ratios of Type 2 Quasars}",
      journal = {\apj},
         year = 2018,
        month = jun,
       volume = {859},
       number = {2},
          eid = {116},
        pages = {116},
          doi = {10.3847/1538-4357/aabe2a},
archivePrefix = {arXiv},
       eprint = {1804.09852},
 primaryClass = {astro-ph.GA},
       adsurl = {https://ui.adsabs.harvard.edu/abs/2018ApJ...859..116K}
}

@ARTICLE{2025AA...698A.194R,
       author = {{Ramos Almeida}, C. and {Garc{\'\i}a-Bernete}, I. and {Pereira-Santaella}, M. and {Speranza}, G. and {Maiolino}, R. and {Ji}, X. and {Audibert}, A. and {Cezar}, P.~H. and {Acosta-Pulido}, J.~A. and {Alonso-Herrero}, A. and {Garc{\'\i}a-Burillo}, S. and {Gonz{\'a}lez-Mart{\'\i}n}, O. and {Rigopoulou}, D. and {Tadhunter}, C.~N. and {Labiano}, A. and {Levenson}, N.~A. and {Donnan}, F.~R.},
        title = "{JWST MIRI reveals the diversity of nuclear mid-infrared spectra of nearby type 2 quasars}",
      journal = {\aap},
         year = 2025,
        month = jun,
       volume = {698},
          eid = {A194},
        pages = {A194},
          doi = {10.1051/0004-6361/202453549},
archivePrefix = {arXiv},
       eprint = {2504.01595},
 primaryClass = {astro-ph.GA},
       adsurl = {https://ui.adsabs.harvard.edu/abs/2025A&A...698A.194R}
}

@ARTICLE{2025arXiv251202629R,
       author = {{Ramos Almeida}, C. and {Asensio Ramos}, A. and {Westerdorp Plaza}, C. and {Garc{\'\i}a-Bernete}, I. and {Lopez-Rodriguez}, E. and {H{\"o}nig}, S. and {Audibert}, A. and {Garc{\'\i}a-Burillo}, S. and {Pereira-Santaella}, M. and {Donnan}, F. and {Alonso-Herrero}, A. and {Gonz{\'a}lez-Mart{\'\i}n}, O. and {Levenson}, N. and {Rigopoulou}, D. and {Tadhunter}, C. and {Speranza}, G.},
        title = "{Silicate emission in a type-2 quasar: JWST/MIRI constraints on torus geometry and radiative feedback}",
      journal = {\aap},
         year = 2026,
        month = mar,
       volume = {706},
          eid = {A100},
        pages = {A100},
          doi = {10.1051/0004-6361/202557323},
archivePrefix = {arXiv},
       eprint = {2512.02629},
 primaryClass = {astro-ph.GA},
       adsurl = {https://ui.adsabs.harvard.edu/abs/2026A&A...706A.100R}
}

@ARTICLE{2021ApJ...916...33G,
       author = {{Gordon}, Karl D. and {Misselt}, Karl A. and {Bouwman}, Jeroen and {Clayton}, Geoffrey C. and {Decleir}, Marjorie and {Hines}, Dean C. and {Pendleton}, Yvonne and {Rieke}, George and {Smith}, J.~D.~T. and {Whittet}, D.~C.~B.},
        title = "{Milky Way Mid-Infrared Spitzer Spectroscopic Extinction Curves: Continuum and Silicate Features}",
      journal = {\apj},
         year = 2021,
        month = jul,
       volume = {916},
          eid = {33},
        pages = {33},
          doi = {10.3847/1538-4357/ac00b7},
archivePrefix = {arXiv},
       eprint = {2105.05087},
 primaryClass = {astro-ph.GA}
}

@ARTICLE{2022ApJS..261....9A,
       author = {{Ananna}, Tonima Tasnim and {Weigel}, Anna K. and {Trakhtenbrot}, Benny and {Koss}, Michael J. and {Urry}, C. Megan and {Ricci}, Claudio and {Hickox}, Ryan C. and {Treister}, Ezequiel and {Bauer}, Franz E. and {Ueda}, Yoshihiro and {Mushotzky}, Richard and {Ricci}, Federica and {Oh}, Kyuseok and {Mej{\'\i}a-Restrepo}, Julian E. and {Brok}, Jakob Den and {Stern}, Daniel and {Powell}, Meredith C. and {Caglar}, Turgay and {Ichikawa}, Kohei and {Wong}, O. Ivy and {Harrison}, Fiona A. and {Schawinski}, Kevin},
        title = "{BASS. XXX. Distribution Functions of DR2 Eddington Ratios, Black Hole Masses, and X-Ray Luminosities}",
      journal = {\apjs},
         year = 2022,
        month = jul,
       volume = {261},
       number = {1},
          eid = {9},
        pages = {9},
          doi = {10.3847/1538-4365/ac5b64},
archivePrefix = {arXiv},
       eprint = {2201.05603},
 primaryClass = {astro-ph.HE},
       adsurl = {https://ui.adsabs.harvard.edu/abs/2022ApJS..261....9A}
}

@ARTICLE{2016MNRAS.458.2288S,
       author = {{Stalevski}, Marko and {Ricci}, Claudio and {Ueda}, Yoshihiro and {Lira}, Paulina and {Fritz}, Jacopo and {Baes}, Maarten},
        title = "{The dust covering factor in active galactic nuclei}",
      journal = {\mnras},
         year = 2016,
        month = may,
       volume = {458},
       number = {3},
        pages = {2288-2302},
          doi = {10.1093/mnras/stw444},
archivePrefix = {arXiv},
       eprint = {1602.06954},
 primaryClass = {astro-ph.GA},
       adsurl = {https://ui.adsabs.harvard.edu/abs/2016MNRAS.458.2288S}
}

@article{Stalevski_2019,
   title={Dissecting the active galactic nucleus in Circinus – II. A thin dusty disc and a polar outflow on parsec scales},
   volume={484},
   ISSN={1365-2966},
   url={http://dx.doi.org/10.1093/mnras/stz220},
   DOI={10.1093/mnras/stz220},
   number={3},
   journal={\mnras},
   publisher={Oxford University Press (OUP)},
   author={Stalevski, Marko and Tristram, Konrad R W and Asmus, Daniel},
   year={2019},
   month=Jan, pages={3334–3355} }

@ARTICLE{miller2025xrismrevealsremnanttorus,
      title={XRISM Reveals a Remnant Torus in the Low-Luminosity AGN M81*}, 
      author={Jon M. Miller and Ehud Behar and Hisamitsu Awaki and Ann Hornschemeier and Jesse Bluem and Luigi Gallo and Shogo B. Kobayashi and Richard Mushotzky and Masanori Ohno and Robert Petre and Kosuke Sato and Yuichi Terashima and Mihoko Yukita},
      year={2025},
      eprint={2505.13730},
      archivePrefix={arXiv},
      primaryClass={astro-ph.HE}, journal={arXiv e-prints}, eid={arXiv:2505.13730},
      url={https://arxiv.org/abs/2505.13730}, 
}

@article{Esparza_Arredondo_2019,
   title={Physical Parameters of the Torus for the Type 2 Seyfert IC 5063 from Mid-IR and X-Ray Simultaneous Spectral Fitting},
   volume={886},
   ISSN={1538-4357},
   url={http://dx.doi.org/10.3847/1538-4357/ab4ced},
   DOI={10.3847/1538-4357/ab4ced},
   number={2},
   journal={\apj},
   publisher={American Astronomical Society},
   author={Esparza-Arredondo, Donaji and González-Martín, Omaira and Dultzin, Deborah and Ramos-Almeida, Cristina and Fritz, Jacopo and Masegosa, Josefa and Pasetto, Alice and Martínez-Paredes, Mariela and Osorio-Clavijo, Natalia and Victoria-Ceballos, Cesar},
   year={2019},
   month=Nov, pages={125} }

@ARTICLE{2021MNRAS.503.5877S,
       author = {{Swain}, Subhashree and {Shalima}, P. and {Latha}, K.~V.~P. and {B S Swamy}, Krishna},
        title = "{Hot graphite dust in the inner regime of NGC 4151}",
      journal = {\mnras},
         year = 2021,
        month = jun,
       volume = {503},
       number = {4},
        pages = {5877-5893},
          doi = {10.1093/mnras/stab372},
archivePrefix = {arXiv},
       eprint = {2102.04662},
 primaryClass = {astro-ph.GA},
       adsurl = {https://ui.adsabs.harvard.edu/abs/2021MNRAS.503.5877S}
}

@ARTICLE{2015ApJ...803..110H,
       author = {{Hatziminaoglou}, E. and {Hern{\'a}n-Caballero}, A. and {Feltre}, A. and {Pi{\~n}ol Ferrer}, N.},
        title = "{A Complete Census of Silicate Features in the Mid-infrared Spectra of Active Galaxies}",
      journal = {\apj},
         year = 2015,
        month = apr,
       volume = {803},
       number = {2},
          eid = {110},
        pages = {110},
          doi = {10.1088/0004-637X/803/2/110},
archivePrefix = {arXiv},
       eprint = {1502.05823},
 primaryClass = {astro-ph.GA},
       adsurl = {https://ui.adsabs.harvard.edu/abs/2015ApJ...803..110H}
}

@ARTICLE{2008MNRAS.391L..49L,
       author = {{Li}, M.~P. and {Shi}, Q.~J. and {Li}, Aigen},
        title = "{On the anomalous silicate emission features of active galactic nuclei: a possible interpretation based on porous dust}",
      journal = {\mnras},
         year = 2008,
        month = nov,
       volume = {391},
       number = {1},
        pages = {L49-L53},
          doi = {10.1111/j.1745-3933.2008.00553.x},
archivePrefix = {arXiv},
       eprint = {0808.4121},
 primaryClass = {astro-ph},
       adsurl = {https://ui.adsabs.harvard.edu/abs/2008MNRAS.391L..49L}
}

@ARTICLE{2017ApJS..228....6X,
       author = {{Xie}, Yanxia and {Li}, Aigen and {Hao}, Lei},
        title = "{Silicate Dust in Active Galactic Nuclei}",
      journal = {\apjs},
         year = 2017,
        month = jan,
       volume = {228},
       number = {1},
          eid = {6},
        pages = {6},
          doi = {10.3847/1538-4365/228/1/6},
archivePrefix = {arXiv},
       eprint = {1612.04293},
 primaryClass = {astro-ph.GA},
       adsurl = {https://ui.adsabs.harvard.edu/abs/2017ApJS..228....6X}
}

@ARTICLE{1993ApJ...402..441L,
       author = {{Laor}, Ari and {Draine}, Bruce T.},
        title = "{Spectroscopic Constraints on the Properties of Dust in Active Galactic Nuclei}",
      journal = {\apj},
         year = 1993,
        month = jan,
       volume = {402},
        pages = {441},
          doi = {10.1086/172149},
       adsurl = {https://ui.adsabs.harvard.edu/abs/1993ApJ...402..441L}
}

@ARTICLE{2020ApJ...892...84T,
       author = {{Tazaki}, Ryo and {Ichikawa}, Kohei and {Kokubo}, Mitsuru},
        title = "{Dust Destruction by Charging: A Possible Origin of Gray Extinction Curves of Active Galactic Nuclei}",
      journal = {\apj},
         year = 2020,
        month = apr,
       volume = {892},
       number = {2},
          eid = {84},
        pages = {84},
          doi = {10.3847/1538-4357/ab7822},
archivePrefix = {arXiv},
       eprint = {2002.08023},
 primaryClass = {astro-ph.GA},
       adsurl = {https://ui.adsabs.harvard.edu/abs/2020ApJ...892...84T}
}

@ARTICLE{2004Natur.429...47J,
       author = {{Jaffe}, W. and {Meisenheimer}, K. and {R{\"o}ttgering}, H.~J.~A. and {Leinert}, Ch. and {Richichi}, A. and {Chesneau}, O. and {Fraix-Burnet}, D. and {Glazenborg-Kluttig}, A. and {Granato}, G.-L. and {Graser}, U. and {Heijligers}, B. and {K{\"o}hler}, R. and {Malbet}, F. and {Miley}, G.~K. and {Paresce}, F. and {Pel}, J.-W. and {Perrin}, G. and {Przygodda}, F. and {Schoeller}, M. and {Sol}, H. and {Waters}, L.~B.~F.~M. and {Weigelt}, G. and {Woillez}, J. and {de Zeeuw}, P.~T.},
        title = "{The central dusty torus in the active nucleus of NGC 1068}",
      journal = {\nat},
         year = 2004,
        month = may,
       volume = {429},
       number = {6987},
        pages = {47-49},
          doi = {10.1038/nature02531},
       adsurl = {https://ui.adsabs.harvard.edu/abs/2004Natur.429...47J}
}

@ARTICLE{2007ApJ...656..770S,
       author = {{Smith}, J.~D.~T. and {Draine}, B.~T. and {Dale}, D.~A. and {Moustakas}, J. and {Kennicutt}, Jr., R.~C. and {Helou}, G. and {Armus}, L. and {Roussel}, H. and {Sheth}, K. and {Bendo}, G.~J. and {Buckalew}, B.~A. and {Calzetti}, D. and {Engelbracht}, C.~W. and {Gordon}, K.~D. and {Hollenbach}, D.~J. and {Li}, A. and {Malhotra}, S. and {Murphy}, E.~J. and {Walter}, F.},
        title = "{The Mid-Infrared Spectrum of Star-forming Galaxies: Global Properties of Polycyclic Aromatic Hydrocarbon Emission}",
      journal = {\apj},
         year = 2007,
        month = feb,
       volume = {656},
       number = {2},
        pages = {770-791},
          doi = {10.1086/510549},
archivePrefix = {arXiv},
       eprint = {astro-ph/0610913},
 primaryClass = {astro-ph},
       adsurl = {https://ui.adsabs.harvard.edu/abs/2007ApJ...656..770S}
}

@ARTICLE{1982MNRAS.199..883B,
  author  = {{Blandford}, R. D. and {Payne}, D. G.},
  title   = {Hydromagnetic flows from accretion discs and the production of radio jets},
  journal = {\mnras},
  year    = {1982},
  volume  = {199},
  pages   = {883--903},
  doi     = {10.1093/mnras/199.4.883}
}

@ARTICLE{2006ApJ...648L.101E,
  author  = {{Elitzur}, M. and {Shlosman}, I.},
  title   = {The AGN-obscuring Torus: The End of the ``Doughnut'' Paradigm?},
  journal = {\apjl},
  year    = {2006},
  volume  = {648},
  pages   = {L101--L104},
  doi     = {10.1086/508158}
}

@ARTICLE{2019ApJ...876..137W,
  author  = {{Williamson}, D. and {Venanzi}, M. and {H{\"o}nig}, S.},
  title   = {3D Radiation Hydrodynamics of a Dynamical Torus},
  journal = {\apj},
  year    = {2019},
  volume  = {876},
  eid     = {137},
  pages   = {137},
  doi     = {10.3847/1538-4357/ab17d5}
}

@ARTICLE{2019ApJ...886...55L,
  author  = {{Leftley}, J. H. and {H{\"o}nig}, S. F. and {Asmus}, D. and
             {Tristram}, K. R. W. and {Gandhi}, P. and {Kishimoto}, M. and
             {Venanzi}, M. and {Williamson}, D. J.},
  title   = {Parsec-scale Dusty Winds in Active Galactic Nuclei: Evidence for Radiation Pressure Driving},
  journal = {\apj},
  year    = {2019},
  volume  = {886},
  eid     = {55},
  pages   = {55},
  doi     = {10.3847/1538-4357/ab4a0b}
}

@ARTICLE{2019MNRAS.489.2177A,
  author  = {{Asmus}, D.},
  title   = {New Evidence for the Ubiquity of Prominent Polar Dust Emission in AGN on Tens of Parsec Scales},
  journal = {\mnras},
  year    = {2019},
  volume  = {489},
  pages   = {2177--2188},
  doi     = {10.1093/mnras/stz2289}
}

@ARTICLE{2020ApJ...893...33L,
  author  = {{L{\'o}pez-Rodr{\'i}guez}, E. and {Alonso-Herrero}, A. and
             {Garc{\'i}a-Burillo}, S. and {Gordon}, M. S. and {Ichikawa}, K. and
             {Imanishi}, M. and {Kameno}, S. and {Levenson}, N. A. and
             {Nikutta}, R. and {Packham}, C.},
  title   = {ALMA Polarimetry Measures Magnetically Aligned Dust Grains in the Torus of NGC 1068},
  journal = {\apj},
  year    = {2020},
  volume  = {893},
  eid     = {33},
  pages   = {33},
  doi     = {10.3847/1538-4357/ab8013}
}

@ARTICLE{2020ApJ...897...26W,
  author  = {{Williamson}, D. and {H{\"o}nig}, S. F. and {Venanzi}, M.},
  title   = {Radiation Hydrodynamics Models of Active Galactic Nuclei: Beyond the Central Parsec},
  journal = {\apj},
  year    = {2020},
  volume  = {897},
  eid     = {26},
  pages   = {26},
  doi     = {10.3847/1538-4357/ab989e}
}

@ARTICLE{2020ApJ...900..174V,
  author  = {{Venanzi}, M. and {H{\"o}nig}, S. F. and {Williamson}, D.},
  title   = {The Role of Infrared Radiation Pressure in Shaping Dusty Winds in AGNs},
  journal = {\apj},
  year    = {2020},
  volume  = {900},
  eid     = {174},
  pages   = {174},
  doi     = {10.3847/1538-4357/aba89f}
}

@article{2010ApJ...716..490S,
  author  = {{Smith}, H. A. and {Li}, A. and {Li}, M. P. and {K{\"o}hler}, M. and
             {Ashby}, M. L. N. and {Fazio}, G. G. and {Huang}, J.-S. and
             {Marengo}, M. and {Wang}, Z. and {Willner}, S. P. and
             {Zezas}, A. and {Spinoglio}, L. and {Wu}, Y. L.},
  title   = {Anomalous Silicate Dust Emission in the Type 1 LINER Nucleus of M81},
  journal = {\apj},
  year    = {2010},
  volume  = {716},
  pages   = {490--503},
  doi     = {10.1088/0004-637X/716/1/490},
  eprint  = {1004.2277},
  archivePrefix = {arXiv}
}

@ARTICLE{2012ApJ...758...66W,
  author  = {{Wada}, K.},
  title   = {Radiation-driven Fountain and Origin of Torus around Active Galactic Nuclei},
  journal = {\apj},
  year    = {2012},
  volume  = {758},
  eid     = {66},
  pages   = {66},
  doi     = {10.1088/0004-637X/758/1/66},
  eprint  = {1208.5272},
  archivePrefix = {arXiv}
}

@article{2015ARA&A..53..365N,
  author  = {{Netzer}, H.},
  title   = {Revisiting the Unified Model of Active Galactic Nuclei},
  journal = {\araa},
  year    = {2015},
  volume  = {53},
  pages   = {365--408},
  doi     = {10.1146/annurev-astro-082214-122302},
  eprint  = {1505.00811},
  archivePrefix = {arXiv}
}

@ARTICLE{2017ApJ...838L..20H,
  author  = {{H{\"o}nig}, S. F. and {Kishimoto}, M.},
  title   = {Dusty Winds in Active Galactic Nuclei: Reconciling Observations with Models},
  journal = {\apjl},
  year    = {2017},
  volume  = {838},
  eid     = {L20},
  pages   = {L20},
  doi     = {10.3847/2041-8213/aa6838},
  eprint  = {1703.07781},
  archivePrefix = {arXiv}
}

@ARTICLE{2017ApJ...843...58C,
  author  = {{Chan}, C.-H. and {Krolik}, J. H.},
  title   = {Geometrically Thick Obscuration by Radiation-driven Outflow from Magnetized Tori of Active Galactic Nuclei},
  journal = {\apj},
  year    = {2017},
  volume  = {843},
  eid     = {58},
  pages   = {58},
  doi     = {10.3847/1538-4357/aa76e4},
  eprint  = {1706.00240},
  archivePrefix = {arXiv}
}

@book{MacKay2003,
  author    = {MacKay, David J. C.},
  title     = {Information Theory, Inference, and Learning Algorithms},
  publisher = {Cambridge University Press},
  address   = {Cambridge},
  year      = {2003}
}

@book{Gregory2005,
  author    = {Gregory, Phil},
  title     = {Bayesian Logical Data Analysis for the Physical Sciences: A Comparative Approach with Mathematica Support},
  publisher = {Cambridge University Press},
  address   = {Cambridge},
  year      = {2005}
}

@inproceedings{Skilling2004,
  author    = {Skilling, John},
  title     = {Nested Sampling},
  booktitle = {Bayesian Inference and Maximum Entropy Methods in Science and Engineering},
  series    = {AIP Conference Proceedings},
  volume    = {735},
  pages     = {395--405},
  year      = {2004},
  doi       = {10.1063/1.1835238}
}

@article{Trotta2008,
  author  = {Trotta, Roberto},
  title   = {Bayes in the sky: Bayesian inference and model selection in cosmology},
  journal = {Contemporary Physics},
  volume  = {49},
  number  = {2},
  pages   = {71--104},
  year    = {2008},
  doi     = {10.1080/00107510802066753}
}

@MISC{Bushouse2024,
  author       = {{Bushouse}, H. and {Eisenhamer}, J. and {Dencheva}, N. and
                  {Davies}, J. and {Greenfield}, P. and {Morrison}, J. and
                  {Hodge}, P. and {Simon}, B. and {Grumm}, D. and others},
  title        = {JWST Calibration Pipeline},
  year         = {2024},
  publisher    = {Zenodo},
  doi          = {10.5281/zenodo.14153298}
}

\begin{appendix}
\onecolumn
\refstepcounter{section}

\begin{sidewaysfigure}[p]
\centering
\includegraphics[width=0.9\textheight]{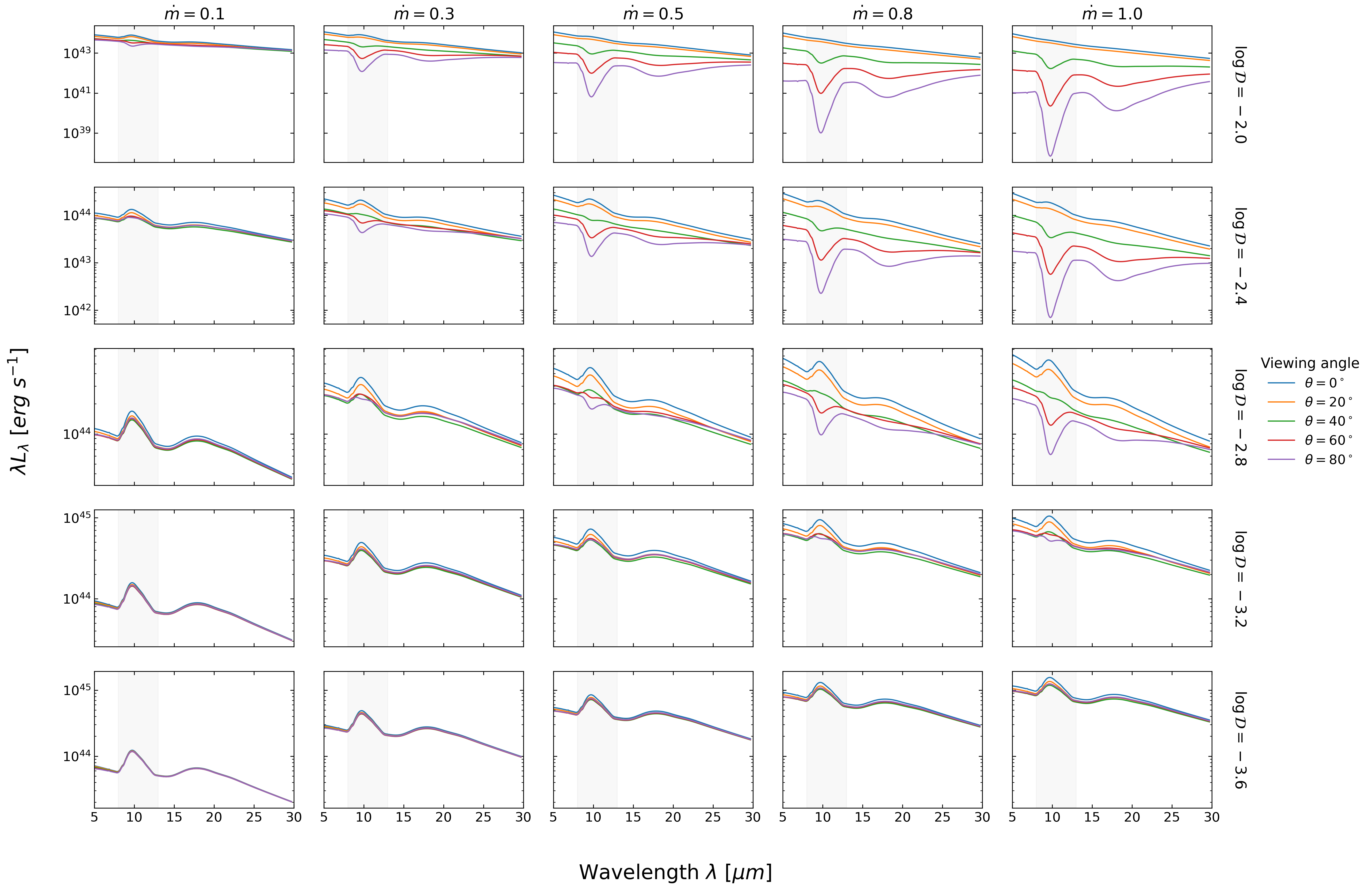}
\caption{Representative grid of rest-frame infrared SEDs ($\lambda L_\lambda$) from the torus library. Panels are organised by normalised accretion rate $\dot{m}$ (columns) and dust-to-gas ratio $\mathcal{D}$ (rows). Different colours indicate viewing angles $\theta_{\rm view}\in[0^\circ,80^\circ]$. All models adopt $M_{\rm BH}=10^{8}\,M_\odot$, $Y=100$, and $\theta_{\rm open}=20^\circ$. The shaded region marks the 8--13~$\mu$m silicate band.}
\label{fig:grid_spectra}
\end{sidewaysfigure}

\clearpage
\begin{table}[p]
\centering
\sisetup{group-digits=false}
\caption{JWST/MRS sample properties}
\label{tab:jwst_sample_extended}
\small
\setlength{\tabcolsep}{3pt}
\renewcommand{\arraystretch}{1.08}
\begin{tabularx}{\textwidth}{@{}l l l S[table-format=1.5] S[table-format=1.3] >{\raggedright\arraybackslash}X >{\raggedright\arraybackslash}p{0.16\textwidth} c@{}}
\toprule
Object & RA (J2000) & Dec (J2000) & {$z$} & {$E(B-V)$} & $M_{\rm BH}$ ($M_\odot$) & {$i_{\rm torus}$} & Optical class\\
\midrule
NGC\,1052 & 02:41:04.8 & $-$08:15:21 & 0.00482 & 0.023 & $1.54\times10^8$ \citep{2009ApJ...698..528B} & 70 \citep{2021ApJ...916...90B} & L1.9\\
ESO\,420-G13 & 04:13:49.7 & $-$30:16:28 & 0.01191 & 0.015 & $4\times10^8$ \citep{2020633A.127F} & / & S2\\
UGC\,05101 & 09:35:51.6 & +61:21:11 & 0.03937 & 0.031 & $1.60\times10^8$ \citep{2022AA...666A.127A} & 65.9 \citep{2021ApJS..257...61Y} & L1.5\\
MCG\,$-$05-23-016 & 09:47:40.1 & $-$30:56:56 & 0.00849 & 0.075 & $1.41\times10^7$ \citep{2012542A..83P} & 50 \citep{2024ApJ...974..195Z} & S2\\
M\,81 & 09:55:33.2 & +69:03:55 & 0.00077 & 0.072 & $7\times10^7$ \citep{2003AJ....125.1226D} & 12 \citep{miller2025xrismrevealsremnanttorus} & S1.8\\
NGC\,3081 & 09:59:29.5 & $-$22:49:35 & 0.00591 & 0.038 & $(0.85$--$3.4)\times10^7$ \citep{2009ApJ...692..856B} & 71 \citep{2024ApJ...974..195Z} & S2\\
NGC\,3256NUC1 & 10:27:51.3 & $-$43:54:14 & 0.00935 & 0.120 & $6.92\times10^6$ \citep{2013ApJ...765...78A} & / & S2\\
NGC\,4395 & 12:25:48.9 & +33:32:48 & 0.00106 & 0.015 & $10^5$--$1.2\times10^6$ \citep{Brok_2015} & 10 \citep{cruz2023modelingsedagninside} & S1.8\\
M\,94 & 12:50:53.1 & +41:07:14 & 0.00103 & 0.014 & $1.6\times10^7$ \citep{2025AA...698L...9N} & / & L2\\
MRK\,231 & 12:56:14.2 & +56:52:25 & 0.04217 & 0.023 & $1.5\times10^8$ \citep{Yan_2015} & 60 \citep{2021ApJS..257...61Y} & S1\\
MRK\,273 & 13:44:42.1 & +55:53:13 & 0.03734 & 0.027 & $1.04\times10^9$ \citep{2013ApJ...775..115U} & 60 \citep{2021ApJS..257...61Y} & S2\\
NGC\,5506 & 14:13:14.9 & $-$03:12:27 & 0.00561 & 0.051 & $8.8\times10^7$ \citep{Niko_ajuk_2009} & 42 \citep{2024ApJ...974..195Z} & S1.9\\
NGC\,5728 & 14:42:23.9 & $-$17:15:11 & 0.00932 & 0.082 & $3.4\times10^7$ \citep{2019ApJ...870...37D} & 49 \citep{2024ApJ...974..195Z} & S2\\
IC\,5063 & 20:52:02.3 & $-$57:04:08 & 0.01135 & 0.055 & $5.5\times10^7$ \citep{Niko_ajuk_2009} & 60 \citep{Esparza_Arredondo_2019} & S2\\
NGC\,7172 & 22:02:01.9 & $-$31:52:11 & 0.00791 & 0.021 & $5.5\times10^7$ \citep{2012ApJ...748..130M} & 67 \citep{2024ApJ...974..195Z} & S2\\
NGC\,7319 & 22:36:03.5 & +33:58:33 & 0.02251 & 0.070 & $2.4\times10^7$ \citep{2022ApJ...936L..14P} & / & S2\\
NGC\,7469 & 23:03:15.6 & +08:52:26 & 0.01627 & 0.060 & $1.22\times10^7$ \citep{2014SSRv..183..253P} & 45 \citep{2021ApJS..257...61Y} & S1.5\\
Circinus & 14:13:09.9 & $-$65:20:21 & 0.00145 & 0.463 & $1.7\times10^6$ \citep{2017ApJ...850...74K} & $>80$ \citep{Stalevski_2019} & S2\\
NGC\,1068 & 02:42:40.7 & $-$00:00:48 & 0.00379 & 0.028 & $1.3\times10^7$ \citep{2020MNRAS.497.1020W} & $>80$ \citep{Gamez-Rosas2023} & S2\\
NGC\,4151 & 12:10:32.6 & +39:24:21 & 0.00332 & 0.023 & $(0.25$--$3)\times10^7$ \citep{2021ApJ...916...25R} & 53 \citep{2021MNRAS.503.5877S} & S1.5\\
J1509+0434 & 15:09:04.2 & +04:34:42 & 0.11150 & 0.041 & $2.00\times10^8$ \citep{2018ApJ...859..116K} & 44 \citep{2025arXiv251202629R} & QSO2\\
J1100+0846 & 11:00:12.4 & +08:46:16 & 0.10040 & 0.025 & $6.31\times10^7$ \citep{2018ApJ...859..116K} & 38 \citep{2025arXiv251202629R} & QSO2\\
J1430+1339 & 14:30:29.9 & +13:39:12 & 0.08510 & 0.031 & $1.58\times10^8$ \citep{2018ApJ...859..116K} & 38 \citep{2025arXiv251202629R} & QSO2\\
J1010+0612 & 10:10:43.3 & +06:12:01 & 0.09770 & 0.022 & $2.51\times10^8$ \citep{2018ApJ...859..116K} & 35 \citep{2025arXiv251202629R} & QSO2\\
J1356+1026 & 13:56:55.7 & +10:25:35 & 0.12320 & 0.019 & $3.98\times10^8$ \citep{2018ApJ...859..116K} & 55 \citep{2025arXiv251202629R} & QSO2\\
\bottomrule
\end{tabularx}
\tablefoot{Sample data were compiled from \citet{2025MNRAS.539.2158G,2025arXiv251202629R}. References for the adopted black hole masses ($M_{\rm BH}$) and torus inclination angles ($i_{\rm torus}$) are listed directly in their respective columns. Unavailable values are denoted by a slash (/). S, L, and QSO2 denote Seyfert, LINER, and type~2 quasar classifications, respectively.}
\end{table}

\clearpage

\begingroup
\setlength{\LTcapwidth}{\linewidth}
\setlength{\tabcolsep}{3pt}
\renewcommand{\arraystretch}{0.95}
\small
\begin{longtable}{@{}l S[table-format=1.1] S[table-format=-3.2] S[table-format=4.1] S[table-format=2.3] c S[table-format=1.1] S[table-format=-1.1] S[table-format=2.0] S[table-format=2.0] S[table-format=3.0] S[table-format=1.3] c@{}}
\caption{\protect\raggedright Fiducial ($\beta=1$) model-comparison results for the $5$--$14\,\mu\mathrm{m}$ AGN+SB fits. For each source, the four rows correspond to radial density laws $n(r)\propto r^{-p}$. The grid score $\log\mathcal{Z}$ is defined in Eq.~(\ref{eq:discrete_evidence}) and is compared only among density laws for the same source. The listed $\chi^2$ values and SB template correspond to the MAP solution, while the other parameters are weighted medians for each density law.}
\label{tab:density_laws_by_object_bayes}\\
\toprule
\multicolumn{1}{c}{Object} & \multicolumn{1}{c}{$p$} & \multicolumn{1}{c}{$\log\mathcal{Z}$} & \multicolumn{1}{c}{$\chi^2_{\rm MAP}$} & \multicolumn{1}{c}{$\chi^2_{\nu,{\rm MAP}}$} & \multicolumn{1}{c}{$M_{\rm BH}/M_\odot$} & \multicolumn{1}{c}{$\dot m$} & \multicolumn{1}{c}{$\log\mathcal{D}$} & \multicolumn{1}{c}{$\theta_{\rm view}$} & \multicolumn{1}{c}{$\theta_{\rm open}$} & \multicolumn{1}{c}{$Y$} & \multicolumn{1}{c}{$f_{\rm AGN}$} & \multicolumn{1}{c}{SB}\\
\midrule
\endfirsthead
\toprule
\multicolumn{1}{c}{Object} & \multicolumn{1}{c}{$p$} & \multicolumn{1}{c}{$\log\mathcal{Z}$} & \multicolumn{1}{c}{$\chi^2_{\rm MAP}$} & \multicolumn{1}{c}{$\chi^2_{\nu,{\rm MAP}}$} & \multicolumn{1}{c}{$M_{\rm BH}/M_\odot$} & \multicolumn{1}{c}{$\dot m$} & \multicolumn{1}{c}{$\log\mathcal{D}$} & \multicolumn{1}{c}{$\theta_{\rm view}$} & \multicolumn{1}{c}{$\theta_{\rm open}$} & \multicolumn{1}{c}{$Y$} & \multicolumn{1}{c}{$f_{\rm AGN}$} & \multicolumn{1}{c}{SB}\\
\midrule
\endhead
\midrule
\multicolumn{13}{r}{\emph{Continued on next page}}\\
\endfoot
\bottomrule
\endlastfoot
NGC\,1052 & 0.5 & 128.61 & 128.3 & 0.844 & $5\times10^8$ & 0.3 & -3.2 & 90 & 20 & 180 & 0.826 & Mrk\,52 \\
 & 1.0 & 77.86 & 234.4 & 1.542 & $10^8$ & 0.1 & -4.0 & 20 & 60 & 180 & 0.647 & NGC\,253 \\
 & 1.5 & 62.11 & 266.3 & 1.752 & $10^8$ & 0.1 & -3.6 & 80 & 40 & 140 & 0.524 & Mrk\,52 \\
 & 2.0 & 41.04 & 307.9 & 2.026 & $10^8$ & 0.1 & -2.8 & 90 & 30 & 100 & 0.505 & Mrk\,52 \\
\midrule
ESO\,420-G13 & 0.5 & 131.69 & 164.1 & 1.080 & $10^8$ & 0.1 & -2.0 & 70 & 50 & 20 & 0.314 & M\,82 \\
 & 1.0 & 130.94 & 167.9 & 1.104 & $5\times10^8$ & 0.1 & -2.0 & 80 & 30 & 60 & 0.296 & M\,82 \\
 & 1.5 & 122.45 & 177.7 & 1.169 & $5\times10^8$ & 0.3 & -2.4 & 90 & 30 & 20 & 0.256 & M\,82 \\
 & 2.0 & 112.23 & 200.6 & 1.320 & $5\times10^8$ & 0.3 & -2.0 & 80 & 20 & 20 & 0.226 & M\,82 \\
\midrule
UGC\,05101 & 0.5 & 46.94 & 368.2 & 2.423 & $5\times10^8$ & 1.0 & -2.0 & 0 & 60 & 180 & 0.330 & IRAS\,17208$-$0014 \\
 & 1.0 & 65.52 & 334.5 & 2.201 & $10^8$ & 1.0 & -2.0 & 30 & 60 & 140 & 0.284 & IRAS\,17208$-$0014 \\
 & 1.5 & 67.00 & 341.0 & 2.244 & $10^8$ & 0.1 & -4.0 & 10 & 60 & 20 & 0.239 & IRAS\,17208$-$0014 \\
 & 2.0 & 79.64 & 327.0 & 2.151 & $5\times10^8$ & 0.1 & -3.6 & 20 & 40 & 100 & 0.241 & IRAS\,17208$-$0014 \\
\midrule
MCG-05-23-016 & 0.5 & 162.16 & 59.6 & 0.392 & $10^7$ & 0.3 & -3.2 & 90 & 30 & 180 & 0.753 & NGC\,253 \\
 & 1.0 & 116.93 & 156.6 & 1.030 & $10^7$ & 0.1 & -2.8 & 80 & 20 & 180 & 0.602 & NGC\,253 \\
 & 1.5 & 86.45 & 218.9 & 1.440 & $10^7$ & 0.1 & -3.2 & 90 & 40 & 100 & 0.534 & NGC\,253 \\
 & 2.0 & 51.76 & 283.4 & 1.864 & $10^7$ & 0.1 & -2.0 & 90 & 20 & 20 & 0.450 & Mrk\,52 \\
\midrule
M\,81 & 0.5 & -5.65 & 422.4 & 2.779 & $5\times10^7$ & 0.1 & -2.8 & 40 & 20 & 180 & 1.000 & Mrk\,52 \\
 & 1.0 & -15.69 & 440.1 & 2.896 & $10^8$ & 0.1 & -4.0 & 0 & 60 & 140 & 0.811 & NGC\,253 \\
 & 1.5 & 0.74 & 406.8 & 2.676 & $5\times10^7$ & 0.1 & -3.6 & 80 & 60 & 140 & 0.685 & Mrk\,52 \\
 & 2.0 & -34.57 & 477.5 & 3.142 & $5\times10^7$ & 0.1 & -2.8 & 90 & 40 & 100 & 0.665 & Mrk\,52 \\
\midrule
NGC\,3081 & 0.5 & 163.04 & 75.7 & 0.498 & $10^7$ & 0.1 & -2.8 & 70 & 20 & 180 & 0.693 & NGC\,253 \\
 & 1.0 & 127.65 & 150.7 & 0.991 & $10^7$ & 0.1 & -3.6 & 70 & 40 & 180 & 0.566 & NGC\,253 \\
 & 1.5 & 92.05 & 220.4 & 1.450 & $10^7$ & 0.1 & -3.2 & 80 & 30 & 100 & 0.434 & Mrk\,52 \\
 & 2.0 & 78.60 & 243.3 & 1.601 & $5\times10^7$ & 0.1 & -2.4 & 90 & 20 & 60 & 0.406 & Mrk\,52 \\
\midrule
NGC\,3256 & 0.5 & 78.40 & 278.0 & 1.829 & $5\times10^6$ & 0.1 & -3.6 & 70 & 30 & 140 & 0.114 & M\,82 \\
 & 1.0 & 67.62 & 296.2 & 1.949 & $10^7$ & 0.1 & -4.0 & 80 & 30 & 140 & 0.094 & M\,82 \\
 & 1.5 & 52.03 & 327.1 & 2.152 & $10^7$ & 0.1 & -3.2 & 90 & 40 & 140 & 0.077 & M\,82 \\
 & 2.0 & 32.59 & 362.1 & 2.382 & $10^7$ & 0.1 & -2.4 & 90 & 20 & 60 & 0.061 & M\,82 \\
\midrule
NGC\,4395 & 0.5 & -504.88 & 1407.5 & 9.260 & $10^6$ & 0.1 & -4.0 & 70 & 20 & 140 & 0.649 & NGC\,253 \\
 & 1.0 & -519.60 & 1435.5 & 9.444 & $10^6$ & 0.1 & -4.0 & 90 & 20 & 140 & 0.494 & Mrk\,52 \\
 & 1.5 & -549.79 & 1495.1 & 9.836 & $10^6$ & 0.1 & -3.2 & 90 & 30 & 140 & 0.432 & Mrk\,52 \\
 & 2.0 & -599.24 & 1591.6 & 10.471 & $10^6$ & 0.1 & -2.4 & 90 & 20 & 100 & 0.375 & Mrk\,52 \\
\midrule
M\,94 & 0.5 & 3.97 & 366.1 & 2.409 & $5\times10^7$ & 0.3 & -2.0 & 0 & 60 & 100 & 0.442 & M\,82 \\
 & 1.0 & 7.81 & 362.2 & 2.383 & $10^7$ & 0.1 & -2.0 & 0 & 60 & 140 & 0.421 & M\,82 \\
 & 1.5 & 23.39 & 333.3 & 2.193 & $10^7$ & 0.1 & -3.6 & 70 & 30 & 140 & 0.388 & M\,82 \\
 & 2.0 & 9.75 & 359.6 & 2.366 & $5\times10^7$ & 0.1 & -2.0 & 80 & 40 & 180 & 0.378 & M\,82 \\
\midrule
MRK\,231 & 0.5 & 188.81 & 15.0 & 0.099 & $5\times10^8$ & 0.1 & -2.4 & 70 & 50 & 100 & 0.735 & Arp\,220 \\
 & 1.0 & 176.34 & 38.0 & 0.250 & $5\times10^8$ & 0.3 & -2.8 & 90 & 30 & 180 & 0.648 & Arp\,220 \\
 & 1.5 & 147.89 & 95.4 & 0.628 & $5\times10^8$ & 0.1 & -2.0 & 80 & 20 & 20 & 0.590 & Arp\,220 \\
 & 2.0 & 133.70 & 124.4 & 0.818 & $10^8$ & 0.3 & -2.4 & 90 & 50 & 20 & 0.501 & NGC\,253 \\
\midrule
MRK\,273 & 0.5 & 140.96 & 161.1 & 1.060 & $10^9$ & 0.3 & -2.0 & 60 & 20 & 180 & 0.384 & IRAS\,17208$-$0014 \\
 & 1.0 & 133.91 & 176.7 & 1.162 & $10^9$ & 0.3 & -2.0 & 70 & 40 & 100 & 0.271 & IRAS\,17208$-$0014 \\
 & 1.5 & 128.33 & 180.7 & 1.189 & $10^9$ & 0.8 & -2.0 & 70 & 40 & 20 & 0.219 & IRAS\,17208$-$0014 \\
 & 2.0 & 128.48 & 189.7 & 1.248 & $5\times10^8$ & 0.3 & -2.4 & 30 & 50 & 100 & 0.163 & IRAS\,17208$-$0014 \\
\midrule
NGC\,5506 & 0.5 & 215.12 & 17.8 & 0.117 & $10^8$ & 0.3 & -2.4 & 60 & 40 & 140 & 0.877 & NGC\,7714 \\
 & 1.0 & 215.05 & 20.1 & 0.132 & $10^8$ & 0.1 & -2.0 & 80 & 20 & 140 & 0.769 & NGC\,253 \\
 & 1.5 & 203.52 & 42.9 & 0.282 & $5\times10^7$ & 0.3 & -2.0 & 90 & 30 & 20 & 0.666 & NGC\,253 \\
 & 2.0 & 189.45 & 66.6 & 0.438 & $10^8$ & 0.3 & -2.0 & 90 & 30 & 20 & 0.613 & NGC\,253 \\
\midrule
NGC\,5728 & 0.5 & 104.77 & 222.3 & 1.463 & $5\times10^7$ & 0.1 & -2.0 & 70 & 20 & 100 & 0.361 & M\,82 \\
 & 1.0 & 104.18 & 222.4 & 1.463 & $10^7$ & 0.8 & -2.0 & 30 & 60 & 140 & 0.184 & M\,82 \\
 & 1.5 & 116.32 & 205.7 & 1.353 & $10^7$ & 0.1 & -4.0 & 20 & 40 & 100 & 0.156 & M\,82 \\
 & 2.0 & 115.43 & 203.5 & 1.339 & $5\times10^7$ & 0.1 & -3.2 & 70 & 40 & 100 & 0.167 & M\,82 \\
\midrule
\pagebreak
IC\,5063 & 0.5 & 149.26 & 96.1 & 0.632 & $5\times10^7$ & 0.1 & -2.8 & 80 & 20 & 180 & 0.631 & NGC\,253 \\
 & 1.0 & 107.69 & 183.6 & 1.208 & $5\times10^7$ & 0.1 & -3.2 & 80 & 30 & 140 & 0.512 & NGC\,253 \\
 & 1.5 & 66.07 & 264.5 & 1.740 & $5\times10^7$ & 0.1 & -3.2 & 90 & 40 & 140 & 0.458 & NGC\,253 \\
 & 2.0 & 36.71 & 320.7 & 2.110 & $5\times10^7$ & 0.1 & -2.4 & 90 & 30 & 20 & 0.371 & Mrk\,52 \\
\midrule
NGC\,7172 & 0.5 & 193.69 & 83.1 & 0.547 & $5\times10^7$ & 0.8 & -2.0 & 50 & 30 & 60 & 0.853 & IRAS\,17208$-$0014 \\
 & 1.0 & 206.47 & 66.1 & 0.435 & $5\times10^7$ & 0.3 & -2.0 & 80 & 40 & 100 & 0.781 & IRAS\,17208$-$0014 \\
 & 1.5 & 194.84 & 83.6 & 0.550 & $5\times10^7$ & 0.5 & -2.0 & 80 & 40 & 20 & 0.673 & IRAS\,17208$-$0014 \\
 & 2.0 & 168.92 & 129.4 & 0.851 & $5\times10^7$ & 1.0 & -2.0 & 80 & 30 & 20 & 0.591 & IRAS\,17208$-$0014 \\
\midrule
NGC\,7319 & 0.5 & 166.76 & 53.7 & 0.353 & $5\times10^7$ & 0.1 & -2.4 & 80 & 20 & 180 & 0.780 & NGC\,253 \\
 & 1.0 & 137.60 & 109.7 & 0.722 & $5\times10^7$ & 0.3 & -2.8 & 90 & 20 & 180 & 0.591 & NGC\,253 \\
 & 1.5 & 119.22 & 149.9 & 0.986 & $5\times10^7$ & 0.1 & -2.0 & 80 & 20 & 20 & 0.522 & NGC\,253 \\
 & 2.0 & 109.08 & 169.4 & 1.114 & $5\times10^7$ & 0.1 & -2.0 & 90 & 30 & 20 & 0.487 & NGC\,253 \\
\midrule
NGC\,7469 & 0.5 & 176.40 & 67.8 & 0.446 & $10^7$ & 0.1 & -2.8 & 80 & 30 & 180 & 0.489 & M\,82 \\
 & 1.0 & 147.54 & 128.0 & 0.842 & $10^7$ & 0.1 & -2.8 & 70 & 20 & 140 & 0.392 & M\,82 \\
 & 1.5 & 130.06 & 160.0 & 1.052 & $5\times10^7$ & 0.1 & -2.0 & 80 & 20 & 20 & 0.318 & NGC\,7714 \\
 & 2.0 & 121.61 & 176.6 & 1.162 & $5\times10^7$ & 0.1 & -2.0 & 90 & 30 & 20 & 0.296 & NGC\,7714 \\
\midrule
Circinus & 0.5 & 130.96 & 123.7 & 0.814 & $5\times10^6$ & 0.1 & -2.4 & 90 & 30 & 180 & 0.455 & NGC\,253 \\
 & 1.0 & 99.77 & 186.3 & 1.226 & $5\times10^6$ & 0.1 & -2.4 & 90 & 30 & 140 & 0.329 & NGC\,253 \\
 & 1.5 & 86.49 & 213.5 & 1.405 & $5\times10^6$ & 0.1 & -2.0 & 90 & 30 & 20 & 0.292 & NGC\,253 \\
 & 2.0 & 73.89 & 238.6 & 1.569 & $10^6$ & 0.1 & -2.0 & 90 & 30 & 20 & 0.266 & NGC\,253 \\
\midrule
NGC\,1068 & 0.5 & 169.46 & 65.3 & 0.430 & $10^7$ & 0.1 & -2.8 & 90 & 30 & 180 & 0.740 & NGC\,253 \\
 & 1.0 & 124.22 & 157.7 & 1.038 & $10^7$ & 0.1 & -2.8 & 80 & 20 & 140 & 0.601 & NGC\,253 \\
 & 1.5 & 89.95 & 226.4 & 1.489 & $10^7$ & 0.1 & -3.2 & 90 & 40 & 100 & 0.533 & NGC\,253 \\
 & 2.0 & 61.56 & 279.9 & 1.842 & $10^7$ & 0.1 & -2.4 & 90 & 30 & 20 & 0.447 & Mrk\,52 \\
\midrule
NGC\,4151 & 0.5 & 194.14 & 23.6 & 0.156 & $10^7$ & 0.1 & -2.4 & 60 & 20 & 180 & 0.982 & Mrk\,52 \\
 & 1.0 & 167.37 & 71.4 & 0.470 & $10^7$ & 0.5 & -3.2 & 70 & 20 & 140 & 0.787 & NGC\,253 \\
 & 1.5 & 146.93 & 117.8 & 0.775 & $10^7$ & 0.1 & -3.2 & 70 & 30 & 60 & 0.687 & NGC\,253 \\
 & 2.0 & 128.88 & 149.5 & 0.983 & $5\times10^7$ & 0.1 & -2.4 & 90 & 20 & 20 & 0.671 & NGC\,253 \\
\midrule
J1509+0434 & 0.5 & 156.49 & 109.4 & 0.720 & $5\times10^8$ & 0.1 & -2.8 & 80 & 50 & 60 & 0.411 & NGC\,7714 \\
 & 1.0 & 148.60 & 126.8 & 0.834 & $5\times10^8$ & 0.1 & -2.4 & 80 & 30 & 100 & 0.337 & NGC\,7714 \\
 & 1.5 & 139.57 & 142.8 & 0.939 & $5\times10^8$ & 0.1 & -2.4 & 80 & 30 & 20 & 0.302 & NGC\,7714 \\
 & 2.0 & 131.82 & 156.2 & 1.028 & $5\times10^8$ & 0.1 & -2.4 & 90 & 30 & 20 & 0.282 & NGC\,7714 \\
\midrule
J1100+0846 & 0.5 & 184.04 & 51.3 & 0.338 & $10^8$ & 0.1 & -2.8 & 80 & 20 & 180 & 0.765 & Mrk\,52 \\
 & 1.0 & 139.92 & 143.9 & 0.946 & $5\times10^7$ & 0.1 & -3.2 & 80 & 20 & 140 & 0.604 & Mrk\,52 \\
 & 1.5 & 109.96 & 202.4 & 1.332 & $10^8$ & 0.1 & -3.2 & 90 & 40 & 100 & 0.529 & Mrk\,52 \\
 & 2.0 & 66.79 & 283.4 & 1.864 & $10^8$ & 0.1 & -2.4 & 90 & 20 & 20 & 0.477 & Mrk\,52 \\
\midrule
J1430+1339 & 0.5 & 171.72 & 48.8 & 0.321 & $5\times10^8$ & 0.1 & -2.4 & 80 & 20 & 140 & 0.707 & Mrk\,52 \\
 & 1.0 & 136.09 & 117.2 & 0.771 & $5\times10^8$ & 0.3 & -2.8 & 90 & 30 & 180 & 0.525 & Mrk\,52 \\
 & 1.5 & 119.20 & 154.3 & 1.015 & $5\times10^8$ & 0.1 & -2.0 & 90 & 30 & 20 & 0.456 & Mrk\,52 \\
 & 2.0 & 105.54 & 181.0 & 1.190 & $5\times10^8$ & 0.1 & -2.0 & 90 & 40 & 20 & 0.409 & Mrk\,52 \\
\midrule
J1010+0612 & 0.5 & 112.98 & 236.6 & 1.557 & $5\times10^8$ & 0.1 & -3.6 & 80 & 60 & 180 & 0.769 & Mrk\,52 \\
 & 1.0 & 33.83 & 392.9 & 2.585 & $10^8$ & 0.1 & -4.0 & 90 & 20 & 180 & 0.640 & Mrk\,52 \\
 & 1.5 & -71.66 & 603.7 & 3.972 & $5\times10^8$ & 0.1 & -3.6 & 90 & 50 & 180 & 0.539 & Mrk\,52 \\
 & 2.0 & -157.38 & 773.1 & 5.086 & $5\times10^8$ & 0.1 & -2.8 & 90 & 20 & 180 & 0.473 & Mrk\,52 \\
\midrule
J1356+1026 & 0.5 & 98.67 & 172.9 & 1.137 & $10^8$ & 0.5 & -3.2 & 90 & 20 & 180 & 0.611 & Arp\,220 \\
 & 1.0 & 66.76 & 245.5 & 1.615 & $10^8$ & 0.1 & -2.0 & 50 & 20 & 140 & 0.414 & NGC\,253 \\
 & 1.5 & 61.42 & 258.0 & 1.698 & $10^8$ & 0.1 & -2.0 & 30 & 20 & 20 & 0.349 & NGC\,253 \\
 & 2.0 & 60.52 & 255.5 & 1.681 & $10^8$ & 0.1 & -2.4 & 80 & 20 & 20 & 0.333 & NGC\,253 \\
\end{longtable}
\endgroup

\begin{figure*}
    \centering
    \includegraphics[width=\textwidth]{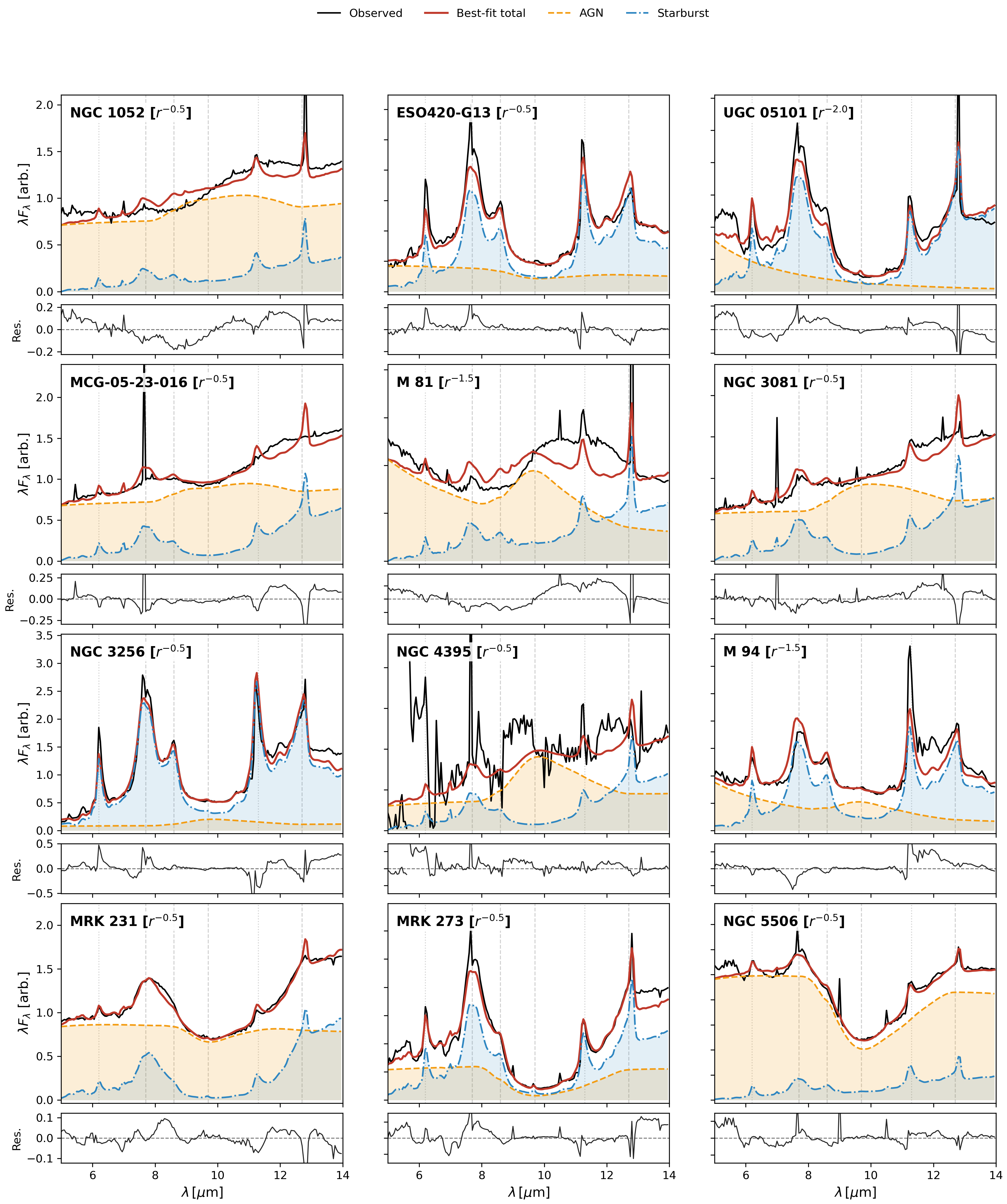}
    \caption{Same as Fig.~\ref{fig:threeexp}, but for the remaining sources. The target name and fiducial ($\beta=1$) preferred density law are given at the upper left of each panel.}
    \label{fig:appendix_spectra}
\end{figure*}

\begin{figure*}
    \addtocounter{figure}{-1}
    \centering
    \includegraphics[width=\textwidth]{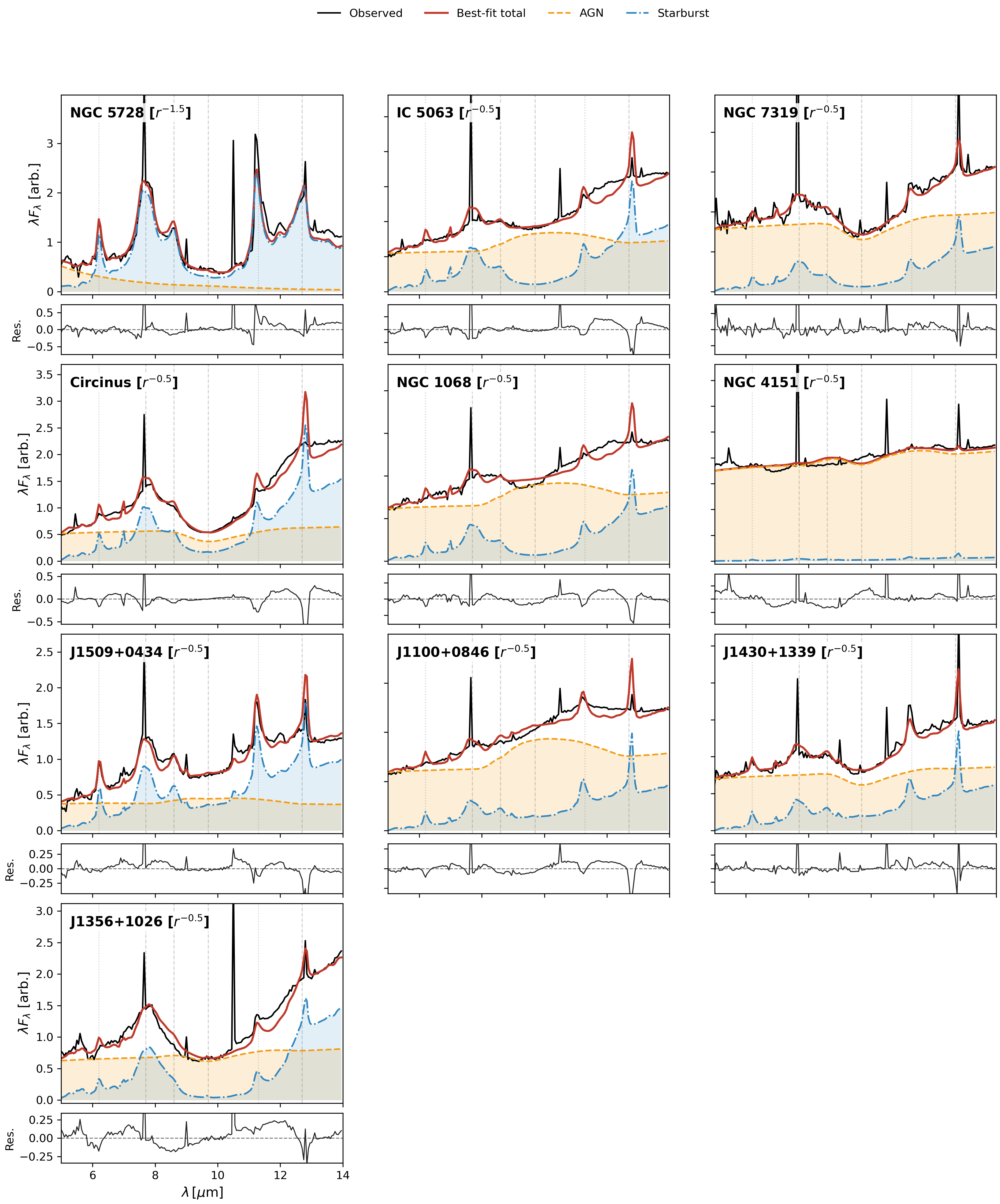}
    \caption{\textit{Continued.}}
\end{figure*}

\end{appendix}

\end{document}